\documentclass[12pt,a4paper]{article}

\usepackage{epsfig}
\usepackage{amsmath,amsfonts,amssymb}
\usepackage{t1enc}
\usepackage{cite}

\providecommand{\openone}{\leavevmode\hbox{\small1\kern-3.8pt\normalsize1}}

\newcommand{\RE}{\text{Re}\,}
\newcommand{\IM}{\text{Im}\,}

\newcommand{\Gp}{\Gamma_+}
\newcommand{\Gm}{\Gamma_-}
\newcommand{\Gz}{\Gamma_0}
\newcommand{\Gpm}{\Gamma_\pm}
\newcommand{\GTp}{\Gamma^T_+}
\newcommand{\GTm}{\Gamma^T_-}
\newcommand{\GTz}{\Gamma^T_0}
\newcommand{\GTpm}{\Gamma^T_\pm}
\newcommand{\GNp}{\Gamma^N_+}
\newcommand{\GNm}{\Gamma^N_-}
\newcommand{\GNz}{\Gamma^N_0}
\newcommand{\GNpm}{\Gamma^N_\pm}
\newcommand{\qb}{|\vec q\,|}
\newcommand{\thlw}{\theta_{\ell}^*}
\newcommand{\thT}{\theta_{\ell}^T}
\newcommand{\thN}{\theta_{\ell}^N}
\newcommand{\fp}{F_+}
\newcommand{\fm}{F_-}
\newcommand{\Fpm}{F_\pm}
\newcommand{\fz}{F_0}
\newcommand{\Fi}{F_i}
\newcommand{\fTp}{F^T_+}
\newcommand{\fTm}{F^T_-}
\newcommand{\fTpm}{F^T_\pm}
\newcommand{\fTz}{F^T_0}
\newcommand{\fNp}{F^N_+}
\newcommand{\fNm}{F^N_-}
\newcommand{\fNpm}{F^N_\pm}
\newcommand{\fNz}{F^N_0}
\newcommand{\thTN}{\theta_{\ell}^{T,N}}
\newcommand{\fxp}{F^{T,N}_+}
\newcommand{\fxm}{F^{T,N}_-}
\newcommand{\fxz}{F^{T,N}_0}
\newcommand{\ftxp}{\tilde F^{T,N}_+}
\newcommand{\ftxm}{\tilde F^{T,N}_-}
\newcommand{\ftxz}{\tilde F^{T,N}_0}
\newcommand{\ftxpm}{\tilde F^{T,N}_\pm}

\newcommand{\afb}{A_\text{FB}}

\newcommand{\afbN}{A_\text{FB}^N}
\newcommand{\afbTN}{A_\text{FB}^{T,N}}
\newcommand{\Ap}{A_+}
\newcommand{\Am}{A_-}

\newcommand{\Apx}{A_+^{T,N}}
\newcommand{\Amx}{A_-^{T,N}}

\newcommand{\all}{A_{\ell \ell}}
\newcommand{\allt}{\tilde A_{\ell \ell}}
\newcommand{\rp}{\rho_+}

\newcommand{\vl}{V_L}
\newcommand{\vr}{V_R}
\newcommand{\gl}{g_L}
\newcommand{\gr}{g_R}

\begin{document}

\begin{center}
\begin{Large}
{\bf $W$ polarisation beyond helicity fractions \\[2mm]
in top quark decays}
\end{Large}
%
% Transverse and normal $W$ polarisation in top decays:
% a new handle to probe the $Wtb$ interaction}
%
% W polarisation beyond helicity fractions in top quark decays
%
% Beyond helicity fractions: transverse and normal W polarisation in top quark decays
%

\vspace{0.5cm}
J. A. Aguilar--Saavedra$^a$, J. Bernab\'eu$^{b,c}$ \\[0.2cm] 
{\it $^a$ Departamento de F\'{\i}sica Te\'orica y del Cosmos and CAFPE, \\
Universidad de Granada, E-18071 Granada, Spain} \\[0.1cm]
{\it $^b$ Departamento de F\'{\i}sica Te\'orica and IFIC, Universidad de Valencia-CSIC, \\ E-46100 Burjassot (Valencia), Spain} \\[0.1cm]
{\it $^c$ CERN, Theory Division, CH-1211 Geneva 23, Switzerland}
\end{center}

\begin{abstract}
We calculate the density matrix for the decay of a polarised top quark into a polarised $W$ boson and a massive $b$ quark, for the most general $Wtb$ vertex arising from dimension-six gauge-invariant effective operators. We show that, in addition to the well-known $W$ helicity fractions, for polarised top decays it is worth defining and studying the transverse and normal $W$ polarisation fractions, that is, the $W$ polarisation along two directions orthogonal to its momentum. In particular, a rather simple forward-backward asymmetry in the normal direction is found to be very sensitive to complex phases in one of the $Wtb$ anomalous couplings. This asymmetry, which indicates a normal $W$ polarisation, can be generated for example by a P-odd, T-odd transition electric dipole moment.
We also investigate the angular distribution of decay products in the top quark rest frame, calculating the spin analysing powers for a general $Wtb$ vertex. Finally we show that, using a combined fit to top decay observables and the $tW$ cross section, at LHC it will be possible to obtain model-independent measurements of all the (complex) $Wtb$ couplings as well as the single top polarisation. Implications for spin correlations in top pair production are also discussed.
\end{abstract}

\section{Introduction}

It is generally believed that the study of the top quark, which is singled out
among the other fermions by its large mass and short lifetime, will be useful to probe
new physics above the electroweak scale~\cite{Beneke:2000hk,Gerber:2007xk, Bernreuther:2008ju,Nath:2010zj}. For this reason, top physics constitutes one of the main programs for Tevatron and the Large Hadron Collider (LHC). Apart from determining the top quark quantum numbers to establish that the top quark is indeed what we expect, its mass and couplings will be
measured. The former is an input parameter in the Lagrangian, whose
precise determination is fundamental to reduce theoretical
uncertainties in many observables. On the other hand, top couplings offer an interesting
window to new physics. If new particles exist above the electroweak scale, their effect at energies below the resonance thresholds can be parameterised by effective operators~\cite{Burges:1983zg,Leung:1984ni,Buchmuller:1985jz} invariant under the standard model (SM) gauge symmetry $\text{SU}(3)_c \times \text{SU}(2)_L \times \text{U}(1)_Y$. In the case of top quark couplings,
the contribution from these operators is expected to be more important
than for the other fermions, due precisely to the large top mass. A general (and minimal) parameterisation of top quark couplings arising from dimension-six effective operators was given in Refs.~\cite{AguilarSaavedra:2008zc}.

Among the different top couplings to the gauge and Higgs bosons, the $Wtb$
vertex deserves a special attention, precisely because the top quark is expected to decay almost exclusively via this interaction, $t \to W b$. Within the effective operator framework, this vertex can be written in full generality as
\begin{eqnarray}
\mathcal{L}_{Wtb} & = & - \frac{g}{\sqrt 2} \bar b \, \gamma^{\mu} \left( \vl
P_L + \vr P_R
\right) t\; W_\mu^- \nonumber \\
& & - \frac{g}{\sqrt 2} \bar b \, \frac{i \sigma^{\mu \nu} q_\nu}{M_W}
\left( \gl P_L + \gr P_R \right) t\; W_\mu^- + \mathrm{h.c.}
\label{ec:lagr}
\end{eqnarray}
This Lagrangian is assumed to be Hermitian in order to preserve unitarity, as it is demanded for a fundamental theory of elementary particle interactions, from which effective operators arise by integration of the heavy degrees of freedom.
This implies that all complex phases in our effective Lagrangian are CP violating. We will not introduce any of the so-called ``CP-conserving phases''~\cite{Antipin:2008zx,Gupta:2009wu,Gupta:2009eq} since they lead to a non-Hermitian Lagrangian with some undesired effects.\footnote{Such phases could appear in the $t \to Wb$ decay amplitude from unitarity corrections, associated with the absorptive parts of higher-order diagrams involving new states lighter than the top quark. The presence of such states, however, contradicts the spirit of the effective operator framework, where new physics is assumed to be heavy, and invariant under the (unbroken) SM gauge group.}

In the SM, the $Wtb$ vertex in Eq.~(\ref{ec:lagr}) reduces to $\vl = V_{tb} \simeq 1$ and $\vr = \gl = \gr = 0$ at the tree level.
Deviations from these values (see for example Refs.~\cite{AguilarSaavedra:2002kr, delAguila:2000aa,Cao:2003yk,Wang:2005ra,Pomarol:2008bh,Bernreuther:2008us}) can be tested by measuring various observables. In particular, the presence of non-zero anomalous couplings $\vr$, $\gl$, $\gr$ is probed with good precision by determining the helicity of the $W$ boson in the top quark decay, {\em i.e.} the relative fractions $\fp$, $\fz$, $\fm$ of $W$ bosons produced with helicity $+1$, $0$, $-1$~\cite{Kane:1991bg}, and through angular distributions in the top quark rest frame~\cite{Jezabek:1994zv,Jezabek:1994qs}. Still, these observables do not contain all the information from the top decay, in particular regarding complex phases. As we will show in this paper, the density matrix for a polarised top quark decay is determined by eight form factors which are functions of the $Wtb$ couplings in Eq.~(\ref{ec:lagr}). Three of these factors appear in the helicity fractions, while the five remaining ones do not. We will find that
a simple and convenient way to probe some of them is by measuring the transverse and normal $W$ polarisation fractions $F_{\pm,0}^T$, $F_{\pm,0}^N$. These are the probabilities for having definite spin components along two directions (transverse and normal) orthogonal to the $W$ momentum.
The normal $W$ polarisation deserves a special mention.
A net normal polarisation ($\fNp \neq \fNm$) unambiguously signals the presence of complex phases in the $Wtb$ vertex because it is directly proportional to the imaginary part of products of $Wtb$ couplings. On the other hand, helicity fractions and distributions in the top quark rest frame involve the real parts of products and moduli squared.
Complex phases can also be probed through triple-product asymmetries in $t \bar t$ production, involving decay products of both $t$ and $\bar t$~\cite{Brandenburg:1992be,Bernreuther:1993hq, Antipin:2008zx,Gupta:2009wu,Gupta:2009eq} but, in contrast, the normal $W$ polarisation can be studied for $t$ (or $\bar t$) decays independently. The power of observables like the transverse and normal polarisations for the study of new physics couplings has been demonstrated for $\tau$ leptons at the $Z$ peak~\cite{Bernabeu:1993er,Alemany:1991ki} and at $B$ factories~\cite{Bernabeu:2006wf,Bernabeu:2007rr}.

The measurement of $W$ transverse and normal polarisation fractions requires the production of polarised top quarks, so that the transverse and normal directions,
defined within and orthogonal to the plane determined by the $W$ momentum and the top polarisation, are meaningful. This will take place, for example, in $t$-channel single top production at LHC, in which the top quarks will have a large polarisation in the direction of the spectator jet~\cite{Mahlon:1999gz}. The determination of $W$ polarisation fractions in this process is expected to achieve a good accuracy, due to the good statistics for this process. In this paper we will show that the measurement of the transverse and normal polarisation fractions (or related observables) will allow to perform a model-independent determination of the complex $Wtb$ vertex in Eq.~(\ref{ec:lagr}), also using helicity fractions, asymmetries in the top quark rest frame and the $tW$ total cross section. A bonus from this analysis is that the single top polarisation, which is taken as a free parameter, can be obtained in a model-independent way from the fit, {\em i.e.} without assumptions on the $Wtb$ couplings. Our fits will be performed with an upgraded version {\tt 2} of the {\tt TopFit} package,\footnote{The code can be downloaded from {\tt http://www-ftae.ugr.es/topfit}.} extended to include many new observables as well as complex anomalous couplings.

We emphasise that
a model-independent determination of the $Wtb$ vertex will be important even if it does not lead to new physics discoveries. Even if new physics does not contribute sizeably to the $Wtb$ vertex and the top quark decays as predicted by the SM, it is crucial to establish this fact in a model-independent way, in order to clearly identify possible new physics in top quark production, if present. One interesting example of this interplay concerns the production of top quark pairs at LHC. Their polarisation (which is very small in the SM) and spin correlation may be modified by the presence of new production mechanisms~\cite{Dicus:1994bm,Bernreuther:1997gs,Choudhury:2007ux}, and hence they probe new physics in $t \bar t$ production. However, the (anti)top polarisation and top-antitop spin correlation can only be measured through angular distributions of $t$, $\bar t$ decay products, which are also sensitive to $Wtb$ anomalous couplings. Performing model-independent measurements of the former obviously requires that the $Wtb$ vertex is precisely measured and possible anomalous couplings are bound. Analogously, in $t$-channel single top production the top polarisation probes new mechanisms for the production, as for example four-fermion operators, new charged gauge bosons and top flavour-changing neutral couplings~\cite{Tait:2000sh,Cao:2007ea,AguilarSaavedra:2010rx}. In this case, a model-independent determination of the single top polarisation (as the one obtained from our fit) is welcome.

The structure of this paper is the following. In the next section we write down the density matrix for polarised (anti)top decays in the helicity basis using the $Wtb$ vertex in Eq.~(\ref{ec:lagr}). In section~\ref{sec:2} we introduce the transverse and normal polarisation fractions, give their expressions, examine their dependence on anomalous couplings and discuss their experimental measurement from angular distributions. Related observables, such as asymmetries in these angular distributions, are defined and studied in section~\ref{sec:3}. In particular, a T-odd forward-backward asymmetry, very sensitive to the phase of $\gr$, is introduced and compared with triple product asymmetries in $t \bar t$ decays. Present and future limits on the transverse and normal polarisation fractions are examined in sections~\ref{sec:4} and \ref{sec:5}, showing that their measurement will bring new information about the $Wtb$ vertex. In section~\ref{sec:6} we present our model-independent fit to the $Wtb$ vertex using estimations for the expected  LHC sensitivities of the different observables. The resulting constraints are used in section~\ref{sec:7} to determine the possible contributions of new physics to the decay vertex and their implications for spin correlations in $t \bar t$ production at LHC. Finally, in section~\ref{sec:8} we adopt the opposite approach: we consider that only one anomalous coupling is non-zero and study the deviations which would show up in the most sensitive observables. We summarise our results in section~\ref{sec:9}. The vector boson polarisation vectors used in our calculations and the relations among them are given in the appendix.

\section{The $t \to Wb$ spin density matrix}
\label{sec:new}

The polarisation of the $W$ bosons produced in the top decay is sensitive
to non-standard $Wtb$ couplings \cite{Kane:1991bg}.
We calculate here the density matrix for the decay of a polarised top quark into a polarised $W$ boson and a  massive $b$ quark. (See Ref.~\cite{Dalitz:1991wa} for an early calculation within the SM.) The program {\tt FORM}~\cite{Vermaseren:2000nd} is used for the symbolic manipulations. For the $W$ boson spin we use the helicity basis (see the appendix) choosing the positive $z$ axis in the direction of its momentum in the top quark rest frame $\vec q$. 
The top spin direction is parameterised as
\begin{equation}
s_t = (0,\sin \theta \cos \phi,\sin \theta \sin \phi,\cos \theta) \,.
\end{equation}
The spin density matrix elements for $W$ helicity components $i,j=0,\pm 1$ are
\begin{equation}
\mathcal{A}(t \to W_i b) \, \mathcal{A}^*(t \to W_j b) = \frac{g^2}{4} m_t^2 M_{ij} \,,
\end{equation}
being
\begin{align}
& M_{00} = A_0 + 2 \frac{\qb}{m_t} A_1 \cos \theta \,, \notag \\
& M_{++} = B_0 \, (1 + \cos \theta) + 2 \frac{\qb}{m_t} B_1 \, (1 + \cos \theta) \,, \notag \\
& M_{--} = B_0 \, (1 - \cos \theta) - 2 \frac{\qb}{m_t} B_1 \, (1 - \cos \theta) \,, \notag \\
& M_{0+} = M_{+0}^* = \left[ \frac{m_t}{\sqrt 2 M_W} (C_0 -i D_0) 
+ \frac{\qb}{\sqrt 2 M_W} (C_1 - i D_1) \right] \sin \theta e^{i \phi} \,, \notag \\
& M_{0-} = M_{-0}^* = \left[ \frac{m_t}{\sqrt 2 M_W} (C_0 -i D_0) 
- \frac{\qb}{\sqrt 2 M_W} (C_1 - i D_1) \right] \sin \theta e^{-i \phi} \,, \notag \\
& M_{+-} = M_{-+} = 0 \,.
\label{ec:Mij}
\end{align}
The dependence on the $Wtb$ couplings in Eq.~(\ref{ec:lagr}) is encoded in eight dimensionless form factors
\begin{align}
A_0 & = \frac{m_t^2}{M_W^2} \left[ |\vl|^2 + |\vr|^2 \right] \left(1 - x_W^2 \right)
+ \left[ |\gl|^2 + |\gr|^2 \right] \left(1 - x_W^2 \right) \notag \\
&  - 4 x_b \, \RE \left[ \vl \vr^* + \gl \gr^* \right]
- 2 \frac{m_t}{M_W} \RE \, \left[\vl \gr^* + \vr \gl^* \right]
\left(1 - x_W^2 \right)  \notag \\
& + 2 \frac{m_t}{M_W} x_b \,\RE \, \left[\vl \gl^* + \vr \gr^* \right]
\left(1 +x_W^2 \right) \,, \notag \\
A_1 & = \frac{m_t^2}{M_W^2} \left[ |\vl|^2 - |\vr|^2 \right]
- \left[ |\gl|^2 - |\gr|^2 \right]
- 2 \frac{m_t}{M_W} \RE \,\left[ \vl \gr^* - \vr \gl^* \right] \notag \\
& + 2 \frac{m_t}{M_W} x_b \RE \, \left[ \vl \gl^* - \vr \gr^* \right] \,, \notag \\
B_0 & = \left[ |\vl|^2 + |\vr|^2 \right] \left(1 - x_W^2 \right) + \frac{m_t^2}{M_W^2} \left[ |\gl|^2 + |\gr|^2 \right] \left(1 - x_W^2 
 \right) \notag \\
& - 4 x_b \, \RE \left[ \vl \vr^* + \gl \gr^* \right]
- 2 \frac{m_t}{M_W} \RE \, \left[\vl \gr^* + \vr \gl^* \right]
\left(1 - x_W^2 \right) \notag \\
& + 2 \frac{m_t}{M_W} x_b \,\RE \, \left[\vl \gl^* + \vr \gr^* \right]
\left(1 +x_W^2 \right) \,, \notag
\displaybreak
 \\
B_1 & = - \left[ |\vl|^2 - |\vr|^2 \right] 
 + \frac{m_t^2}{M_W^2} \left[ |\gl|^2 - |\gr|^2 \right]
+ 2 \frac{m_t}{M_W} \, \RE \, \left[\vl \gr^* - \vr \gl^* \right]  \notag \\
&  + 2 \frac{m_t}{M_W} x_b \, \RE \, \left[\vl \gl^* - \vr \gr^* \right] \,,
\notag \\
C_0 & = \left[ |\vl|^2 + |\vr|^2 + |\gl|^2 + |\gr|^2  \right] \left(1 - x_W^2 \right) 
- 2 x_b \, \RE \left[ \vl \vr^* + \gl \gr^* \right] \left( 1+x_W^2 \right) \notag \\
& - \frac{m_t}{M_W} \RE \, \left[\vl \gr^* + \vr \gl^* \right]
\left(1 - x_W^4 \right) + 4 x_W x_b \,\RE \, \left[\vl \gl^* + \vr \gr^* \right]
\,, \notag \\
C_1 & = 2 \left[ -|\vl|^2 + |\vr|^2 + |\gl|^2 - |\gr|^2 \right] + 2 \frac{m_t}{M_W}
\RE \, \left[ \vl \gr^* - \vr \gl^* \right] \left( 1+x_W^2 \right)
\,, \notag \\
D_0 & = \frac{m_t}{M_W} \IM \, \left[ \vl \gr^* + \vr \gl^* \right] 
\left( 1-2 x_W^2 + x_W^4 \right) \,, \notag \\
D_1 & = -4 x_b \, \IM \left[ \vl \vr^* + \gl \gr^* \right] -2 \frac{m_t}{M_W}
\IM \left[\vl \gr^*-\vr \gl^* \right] (1-x_W^2) \,,
\label{ec:AtoF}
\end{align}
with $x_W = M_W/m_t$, $x_b = m_b/m_t$. The $W$ momentum in the top quark rest frame is
\begin{equation}
\qb = \frac{m_t}{2} (1 - x_W^2) \,.
\end{equation}
We emphasise that, while in the SM it is safe to neglect the $b$ quark mass~\cite{Kane:1991bg,Chen:2005vr}, in the presence of the anomalous couplings $\vr$ and $\gl$ this is no longer possible~\cite{Espriu:2001vj,AguilarSaavedra:2006fy}. Indeed, linear interference terms like $x_b \, \vl \gl^*$ and $x_b \, \vl \vr^*$ can be of the same size as the quadratic ones $|\gl|^2$, $|\vr|^2$ for $\gl$ and $\vr$ small. On the other hand, in the above expressions we have omitted terms of order $x_b^2$ and higher, which amount to corrections of the order of $10^{-3}$ or smaller. All terms are kept in our numerical code, however. The best sensitivity is expected for both $\RE \gr$ and $\IM \gr$, due to their interference with $\vl$ without any suppression by $x_b$.
It is also worthwhile to remark here that $D_0$, $D_1$ are proportional to the imaginary parts of products of $Wtb$ couplings, in contrast with the other terms which contain the moduli squared and the real parts. The form factors $D_0$, $D_1$ are thus entirely new physics effects.
The spin density matrix elements $\bar M_{ij}$ for antitop decays are
\begin{align}
& \bar M_{00} = A_0 - 2 \frac{\qb}{m_t} A_1 \cos \theta \,, \notag \\
& \bar M_{++} = B_0 \, (1 + \cos \theta) - 2 \frac{\qb}{m_t} B_1 \, (1 + \cos \theta) \,, \notag \\
& \bar M_{--} = B_0 \, (1 - \cos \theta) + 2 \frac{\qb}{m_t} B_1 \, (1 - \cos \theta) \,, \notag \\
& \bar M_{0+} = \bar M_{+0}^* = \left[ \frac{m_t}{\sqrt 2 M_W} (C_0 +i D_0) 
- \frac{\qb}{\sqrt 2 M_W} (C_1 + i D_1) \right] \sin \theta e^{i \phi} \,, \notag \\
& \bar M_{0-} = \bar M_{-0}^* = \left[ \frac{m_t}{\sqrt 2 M_W} (C_0 +i D_0) 
+ \frac{\qb}{\sqrt 2 M_W} (C_1 + i D_1) \right] \sin \theta e^{-i \phi} \,, \notag \\
& \bar M_{+-} = \bar M_{-+} = 0 \,.
\label{ec:MBij}
\end{align}

\section{$W$ polarisation beyond helicity fractions}
\label{sec:2}

The partial widths for the top decay into a $W$ boson with $+1$, $0$ or $-1$ helicity, denoted here as $\Gp$, $\Gz$, $\Gm$ respectively, can be straightforwardly obtained from Eqs.~(\ref{ec:Mij}), (\ref{ec:AtoF}) by integrating over $\cos \theta$, $\phi$ and including the appropriate phase space factors. They are~\cite{AguilarSaavedra:2006fy}
\begin{align}
& \Gz = \frac{g^2 \qb}{32 \pi} A_0 \,, && \Gpm = \frac{g^2 \qb}{32 \pi} \left(B_0 \pm 2 \frac{\qb}{m_t} B_1 \right)  \,.
\end{align}
Since the total width $\Gamma(t \to Wb) = \Gm + \Gz + \Gp$ is about 8
times smaller than the expected width of the top invariant mass
peak~\cite{Aad:2009wy,Ball:2007zza}, measuring deviations in $\Gamma$ due to
anomalous couplings or $\vl$ different from one seems rather difficult.
Instead, the $W$ helicity fractions $\Fi \equiv \Gamma_i/\Gamma$
are usually studied. At the tree level,
$\fm = 0.2971$, $\fz = 0.7025$, $\fp = 0.000359$ in the SM for $m_t = 175$ GeV, $M_W = 80.4$ GeV, $m_b = 4.8$ GeV. At NNLO in QCD, $\fm = 0.311$, $\fz = 0.687$, $\fp = 0.0017$~\cite{Czarnecki:2010gb} for a slightly smaller value of the top quark mass $m_t=172.8$ GeV.

Helicity fractions can be measured in leptonic decays $W \to \ell \nu$. Let us denote by $\thlw$ the angle between the charged lepton
three-momentum in the $W$ rest frame and the $W$ momentum in the $t$ rest frame (corresponding to the spin axis in the helicity basis). Then, the normalised angular distribution of the charged lepton is given by
\begin{equation}
\frac{1}{\Gamma} \frac{d \Gamma}{d\!\cos \thlw} = \frac{3}{8}
(1 + \cos \thlw)^2 \, \fp + \frac{3}{8} (1-\cos \thlw)^2 \, \fm
+ \frac{3}{4} \sin^2 \thlw \, \fz \,,
\label{ec:dist}
\end{equation}
with the three terms corresponding to the three helicity states.\footnote{Note that the off-diagonal terms of the spin density matrix give vanishing integral, which implies
that $\sum_i |\mathcal{A}(t \to W_i b \to \ell \nu)|^2 \propto \sum_i |\mathcal{A}(t \to W_i b)|^2 \times |\mathcal{A}(W_i \to \ell \nu)|^2$ in the narrow width approximation and justifies the use of Eq.~(\ref{ec:dist}) for this basis. Moreover, the off-diagonal terms in the $W \to \ell \nu$ density matrix vanish when integrated on the azimuthal angle with respect to the $W$ spin quantisation axis, which also justifies this decomposition for any basis.}
A fit to the $\cos \thlw$ distribution allows to extract from experiment the values of $\Fi$, which are not independent but satisfy $\fp + \fm + \fz = 1$ by definition.

\begin{figure}[htb]
\begin{center}
\epsfig{file=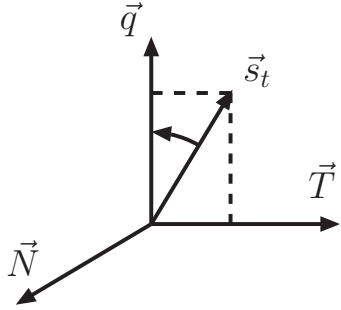,height=4cm,clip=}
\caption{Spin axes defined for the decay of a polarised top quark.}
\label{fig:axes}
\end{center}
\end{figure}

For unpolarised top quark decays, the only meaningful direction in the top quark rest frame is the one of the $W$ boson (and $b$ quark) three-momentum. However,
for polarised top quark decays further spin directions may be considered, as indicated in Fig.~\ref{fig:axes}:
\begin{itemize}
\item[(i)] the transverse direction $\vec T$, defined as the axis orthogonal to the $W$ momentum $\vec q$ and contained in the plane defined by it and the top quark spin direction $\vec s_t$,
\item[(ii)] the normal direction $\vec N$, perpendicular to the plane defined by the $W$ momentum and the top spin direction.
\end{itemize}

We define the transverse and normal vectors as
\begin{align}
& \vec N = \vec s_t \times \vec q \,, \notag \\
& \vec T = \vec q \times \vec N \,,
\label{ec:dir}
\end{align}
corresponding to the ones shown in the figure. For these two spin directions, two further sets of polarised $W$ partial widths can be defined, $\GTm$, $\GTz$, $\GTp$ (transverse) and $\GNm$, $\GNz$, $\GNp$ (normal). They can be obtained either (i) by direct computation 
using the polarisation vectors given in the appendix, or (ii) by using their relation with the helicity basis and the spin density matrix elements for $\phi=\pi/2$. We have performed both calculations as a cross-check. The polarised partial widths for a general $Wtb$ vertex are
\begin{align}
& \GTz = \GNz = \frac{g^2 \qb}{32 \pi} B_0 \,,
&& \GTpm = \frac{g^2 \qb}{32 \pi} \left(\frac{A_0+B_0}{2} \pm \frac{\pi}{4} \frac{m_t}{M_W} C_0 \right) \,,
\notag \\
& && \GNpm = \frac{g^2 \qb}{32 \pi} \left(\frac{A_0+B_0}{2} \pm \frac{\pi}{4} \frac{\qb}{M_W} D_1 \right) \,.
\end{align}
These quantities are very useful to access some of the off-diagonal terms in the spin density matrix, namely $C_0$ and $D_1$.
We point out that $\GNm = \GNp$ if CP is conserved in the $Wtb$ vertex, {\em i.e.} if all anomalous couplings are real ($\vl$ can always be made real with a redefinition of the quark fields). This implies that a net normal $W$ polarisation ($\GNm \neq \GNp$) can only be produced if CP is violated in the $t \to Wb$ decay.\footnote{We note that the normal polarisation $\GNp-\GNm$ is T-odd but not a genuine CP-violating observable, if absorptive parts were present in the decay amplitude.}
This property is unique to the normal direction. Although the helicity and transverse polarisation (as well as top rest frame distributions, see section~\ref{sec:5}) obviously depend quadratically on the imaginary part of anomalous couplings through the moduli squared, their measurement cannot clearly signal the presence of complex phases in the $Wtb$ vertex as the normal polarisation can, through the linear interference term $\IM \vl \gr^*$.

\begin{figure}[htb]
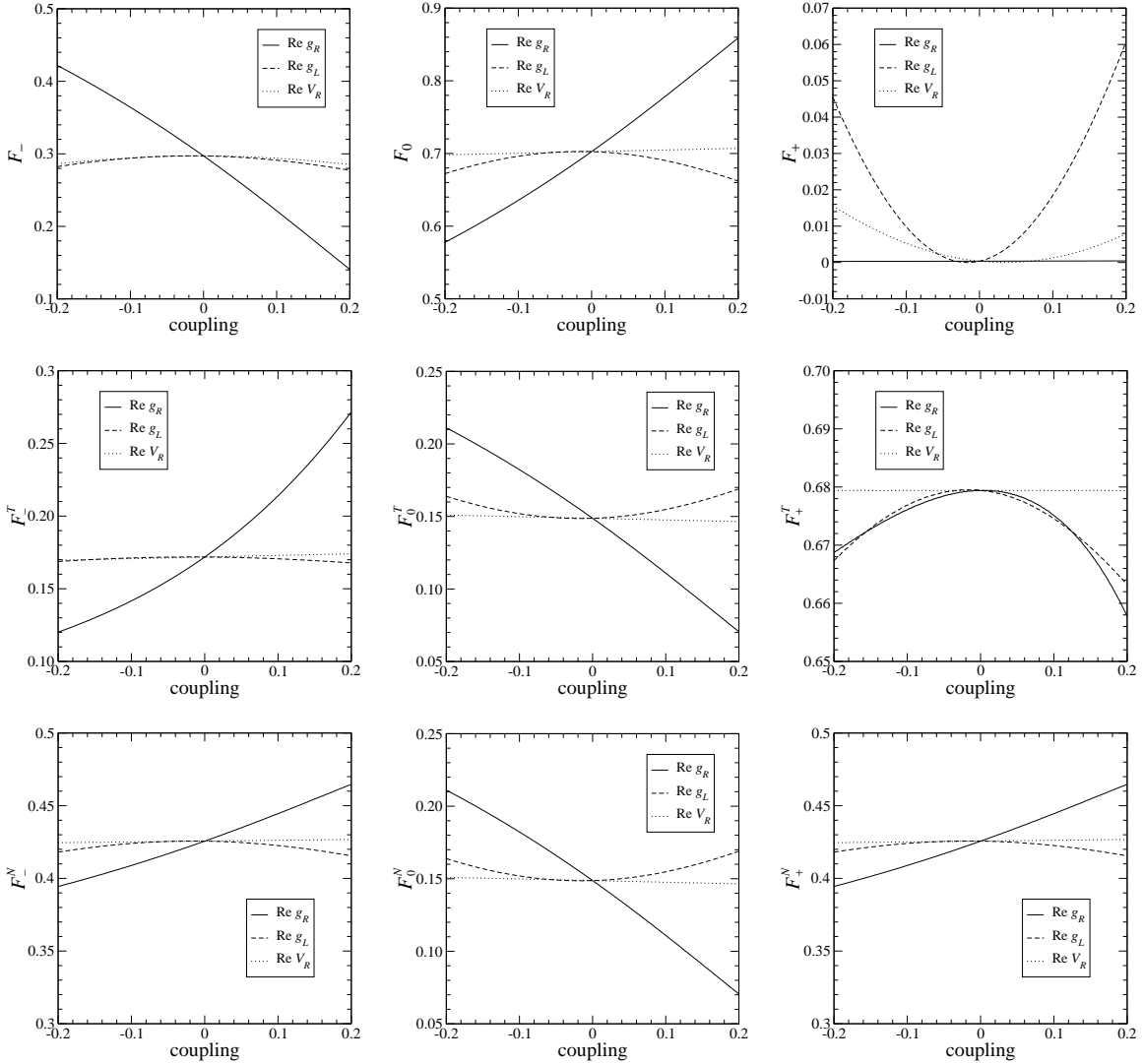

\begin{center}
\begin{tabular}{ccc}
\epsfig{file=Figs/FL.eps,height=4.5cm,clip=} &
\epsfig{file=Figs/F0.eps,height=4.5cm,clip=} &
\epsfig{file=Figs/FR.eps,height=4.5cm,clip=} \\[2mm]
\epsfig{file=Figs/FTL.eps,height=4.5cm,clip=} &
\epsfig{file=Figs/FT0.eps,height=4.5cm,clip=} &
\epsfig{file=Figs/FTR.eps,height=4.5cm,clip=} \\[2mm]
\epsfig{file=Figs/FNL.eps,height=4.5cm,clip=} &
\epsfig{file=Figs/FN0.eps,height=4.5cm,clip=} &
\epsfig{file=Figs/FNR.eps,height=4.5cm,clip=}
\end{tabular}
\caption{Dependence of the $W$ polarisation fractions on the real part of $Wtb$ anomalous couplings in Eq.~(\ref{ec:lagr}), taking $\vl = 1$ and imaginary parts vanishing. Up, middle, down: longitudinal, transverse and normal polarisation fractions, respectively.}
\label{fig:F}
\end{center}
\end{figure}

\begin{figure}[htb]
\begin{center}
\begin{tabular}{ccc}
\epsfig{file=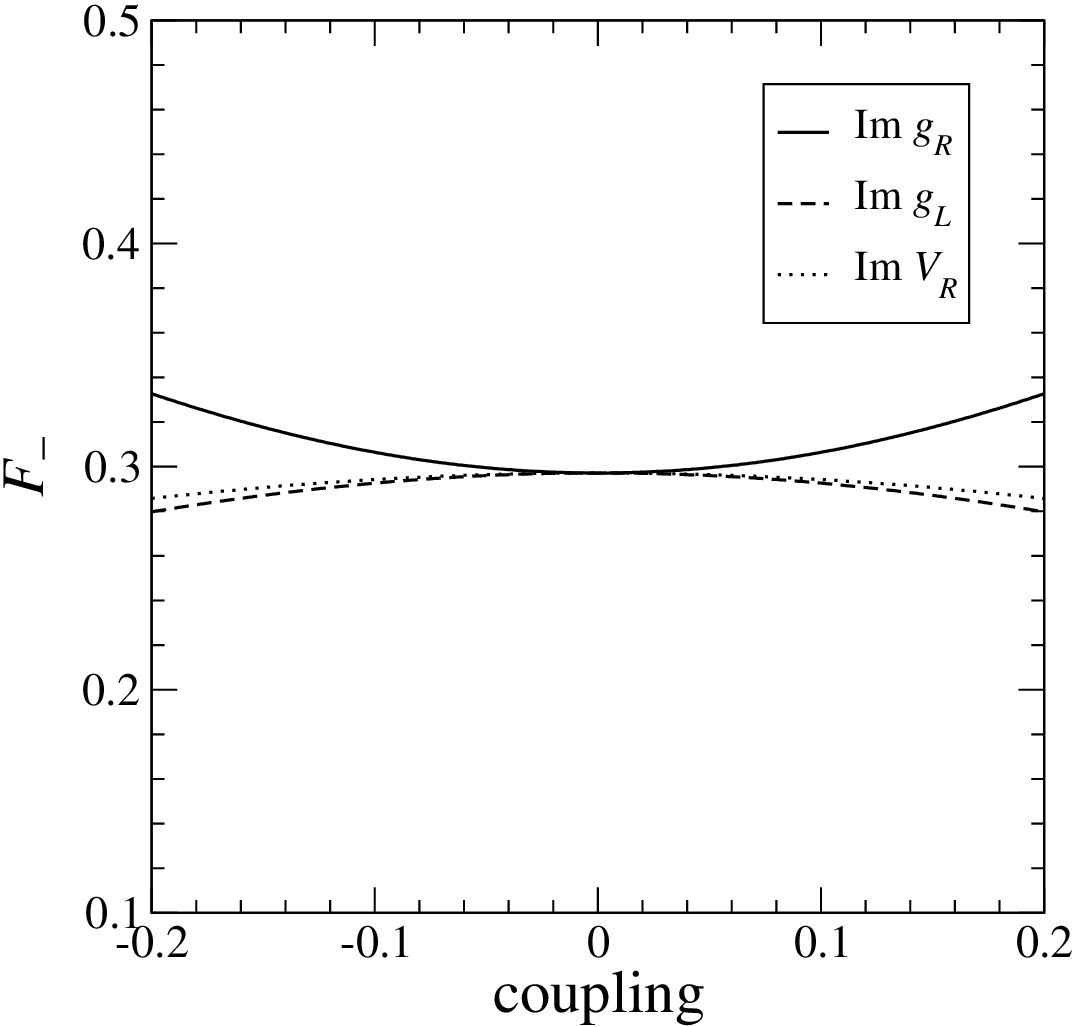,height=4.5cm,clip=} &
\epsfig{file=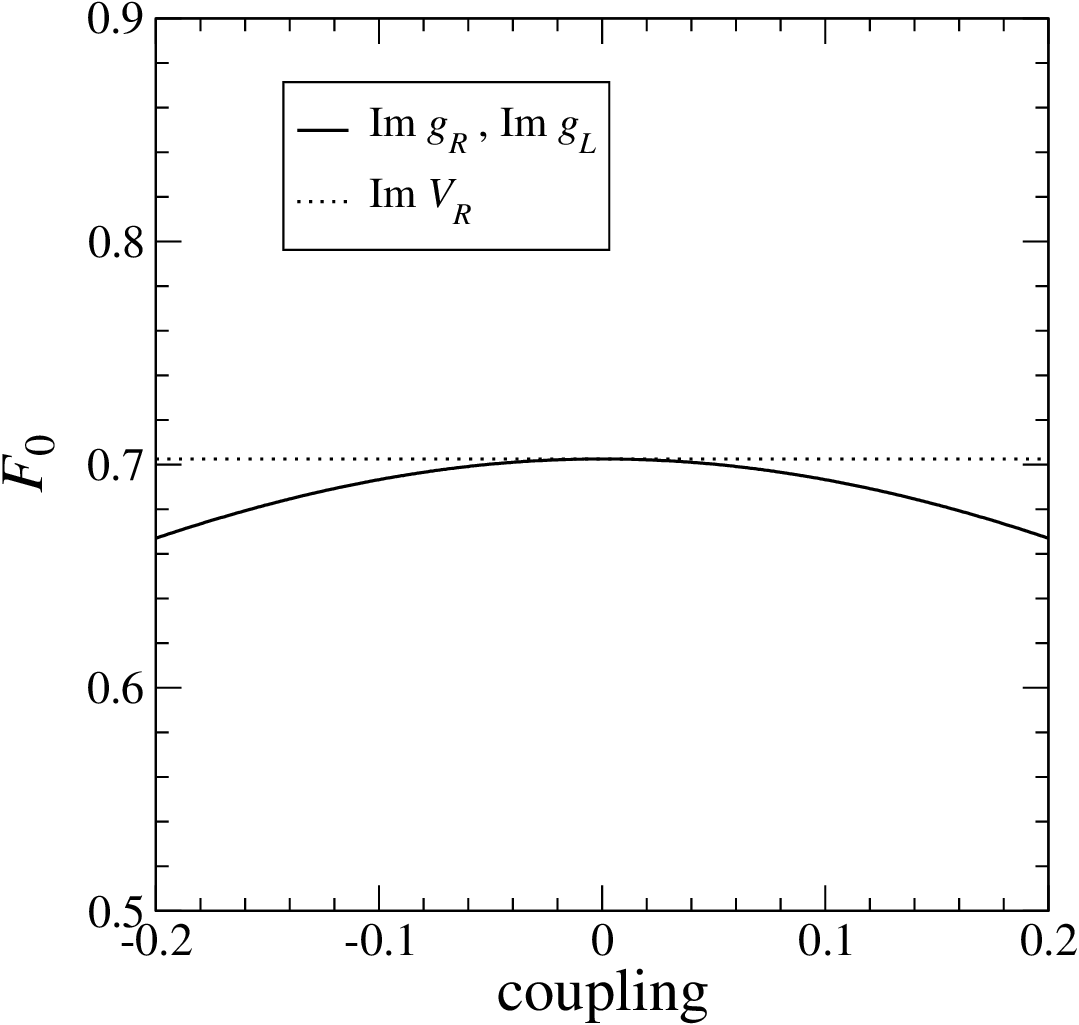,height=4.5cm,clip=} &
\epsfig{file=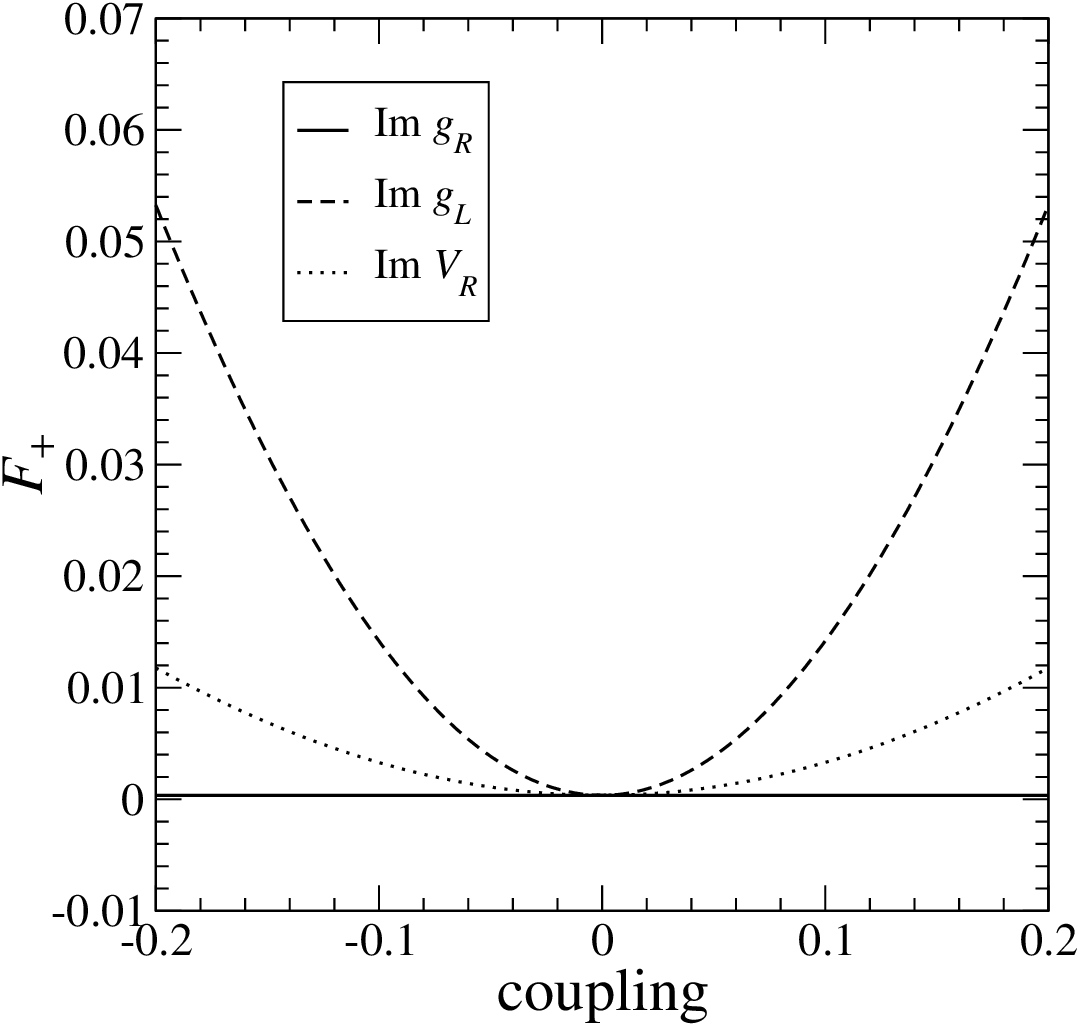,height=4.5cm,clip=} \\[2mm]
\epsfig{file=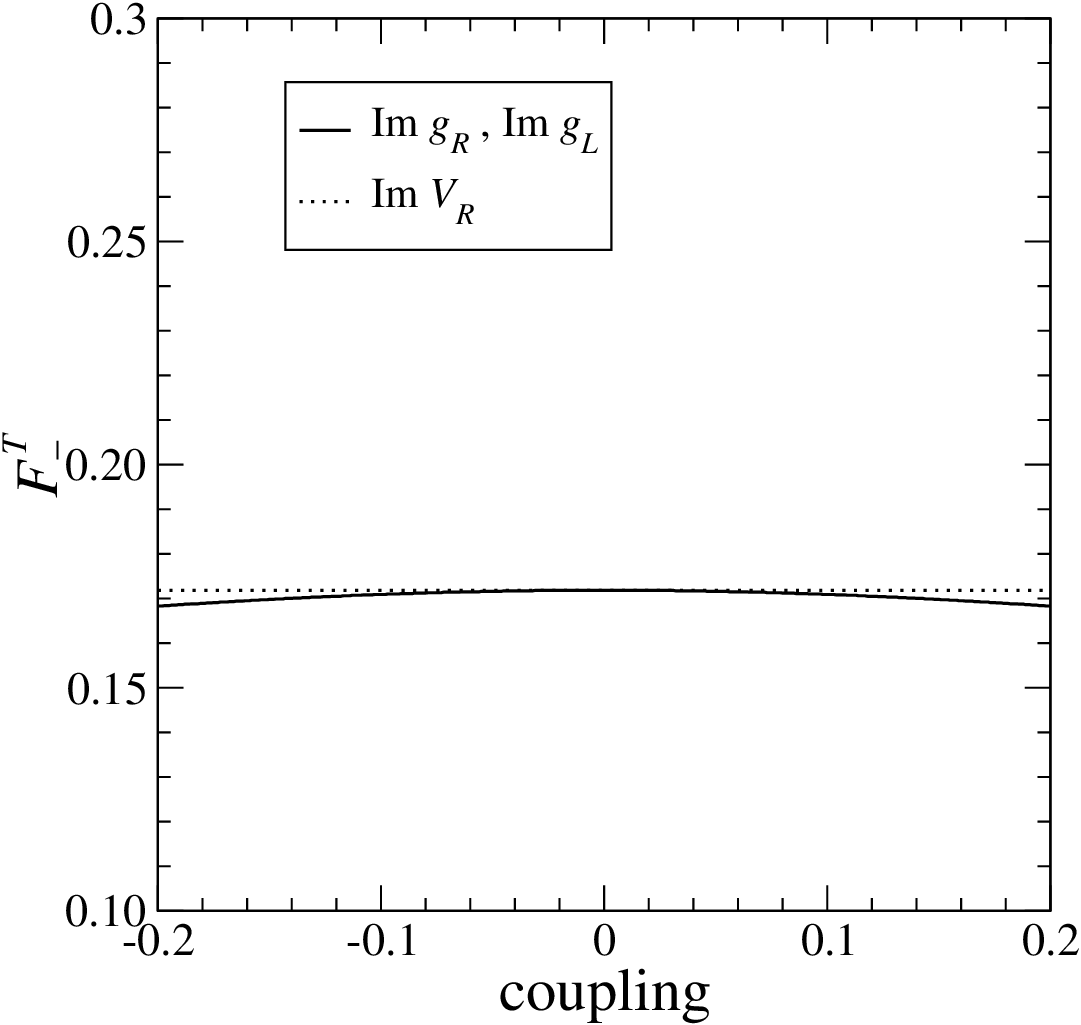,height=4.5cm,clip=} &
\epsfig{file=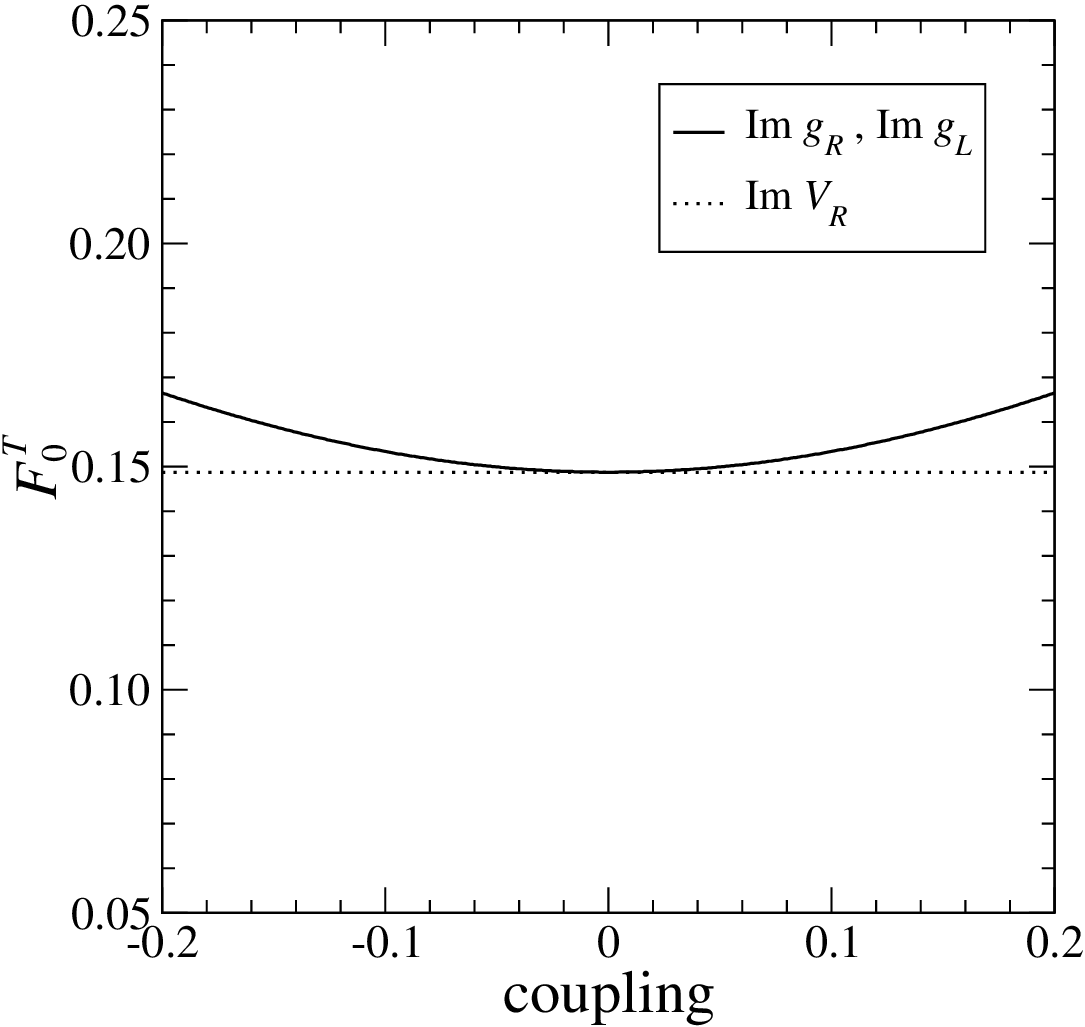,height=4.5cm,clip=} &
\epsfig{file=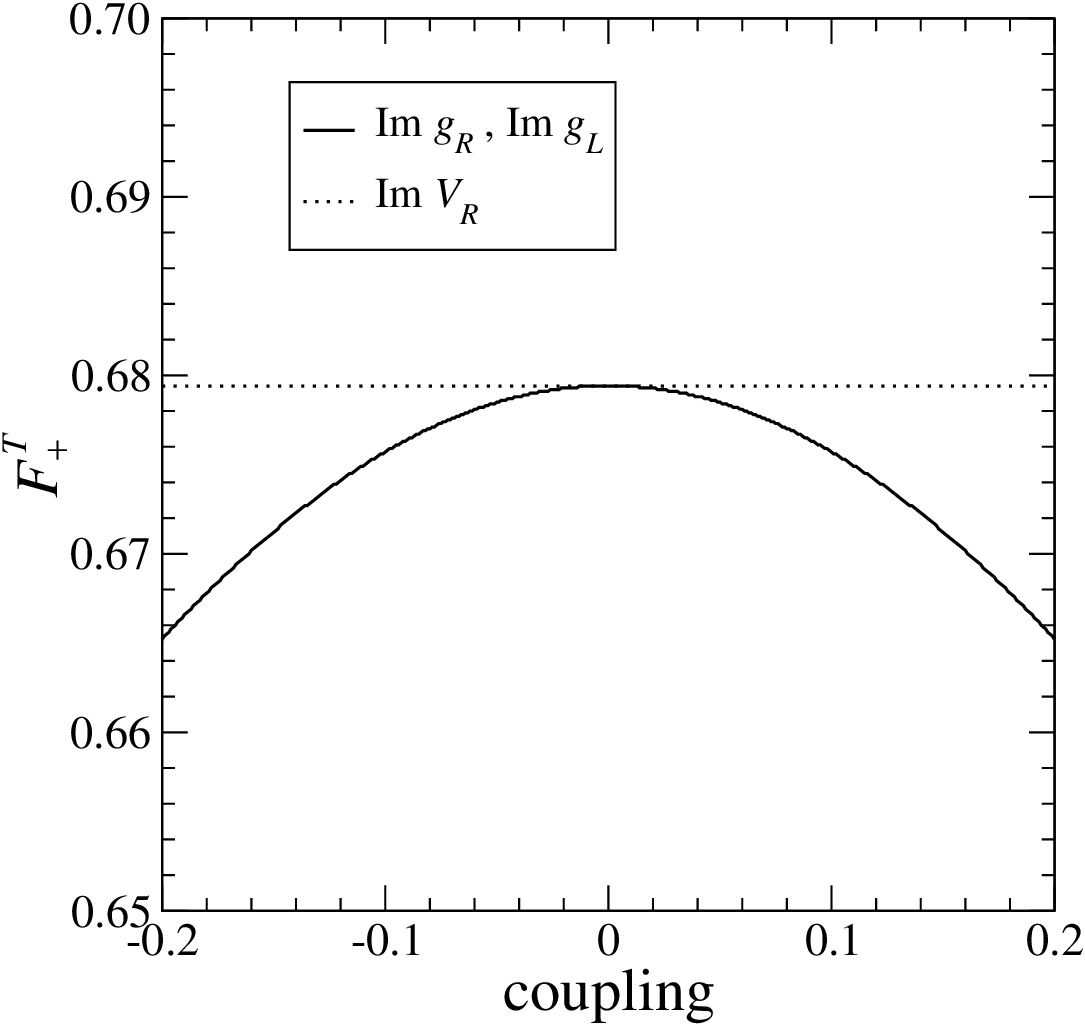,height=4.5cm,clip=} \\[2mm]
\epsfig{file=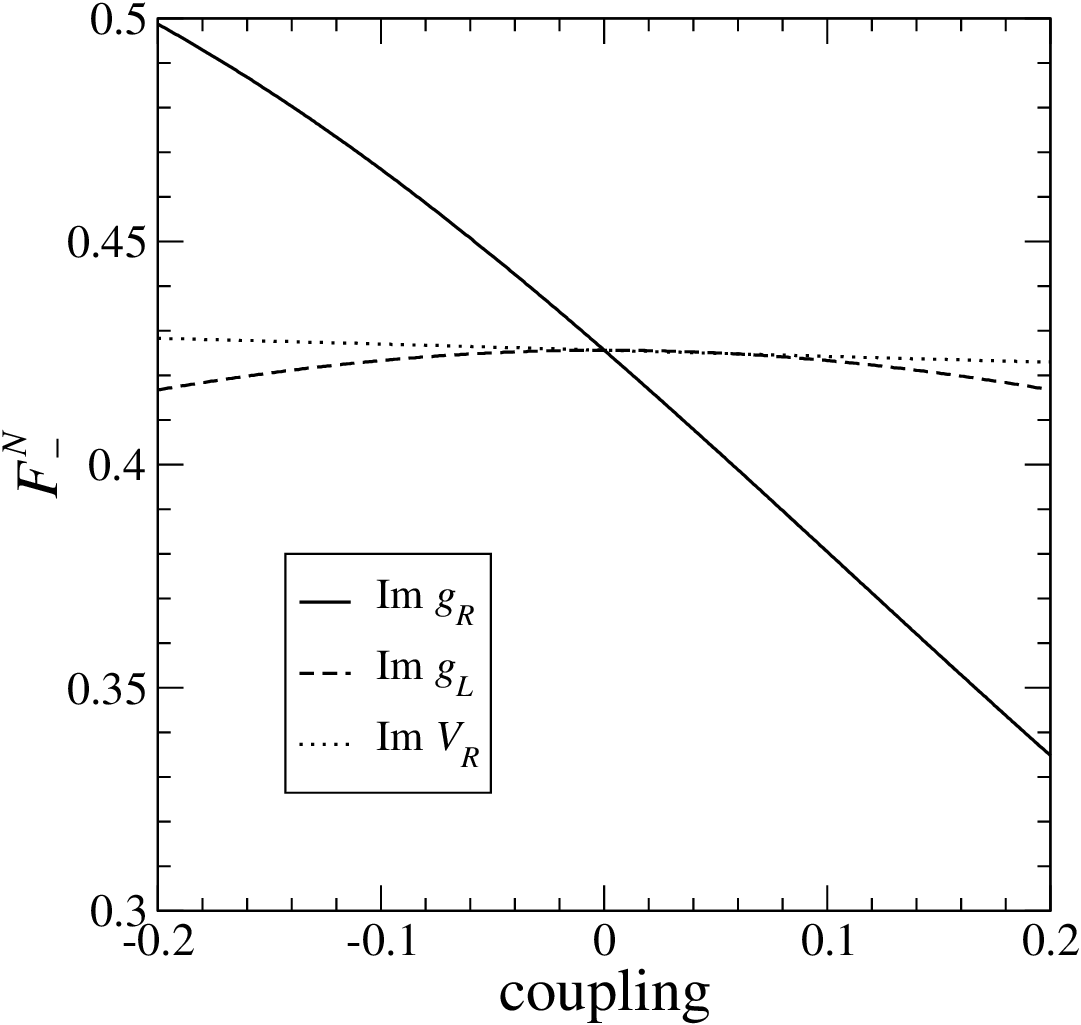,height=4.5cm,clip=} &
\epsfig{file=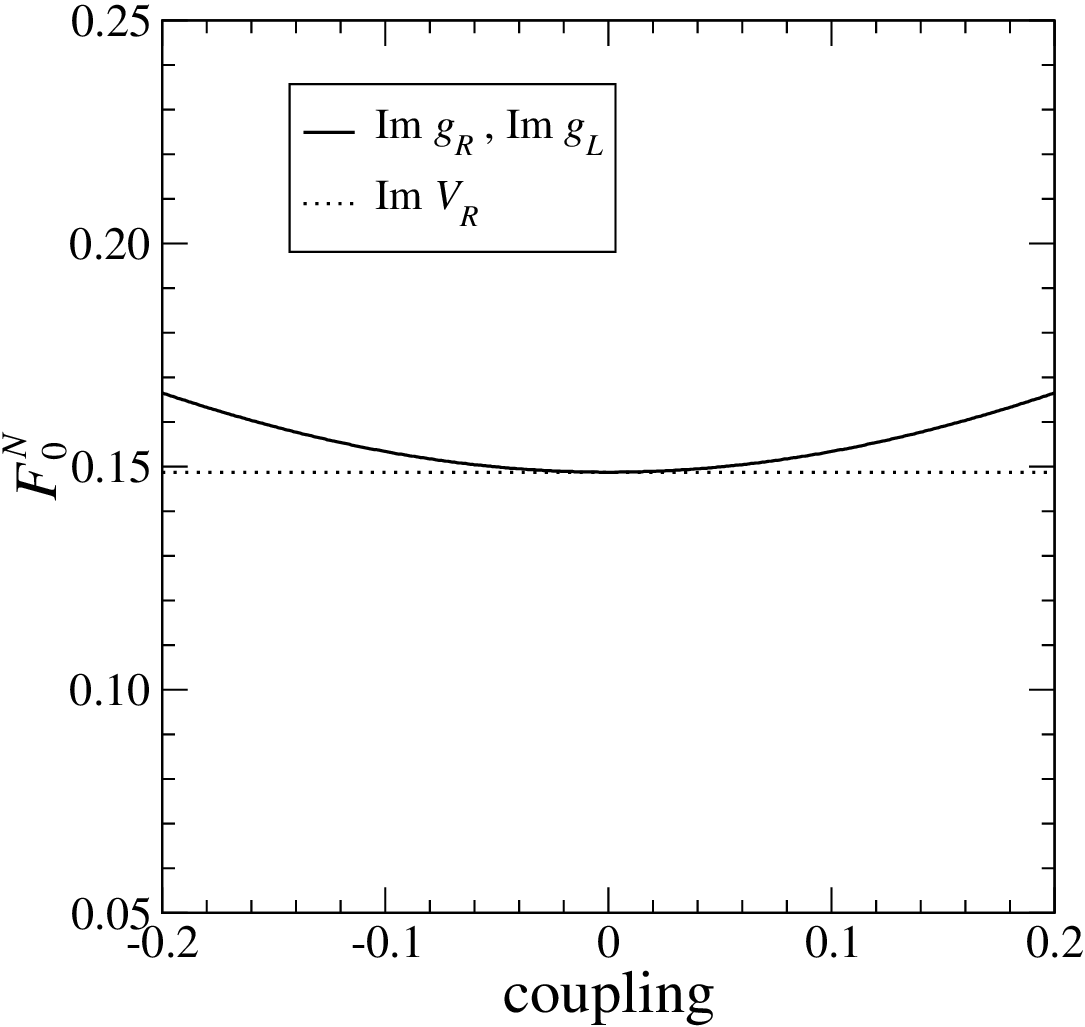,height=4.5cm,clip=} &
\epsfig{file=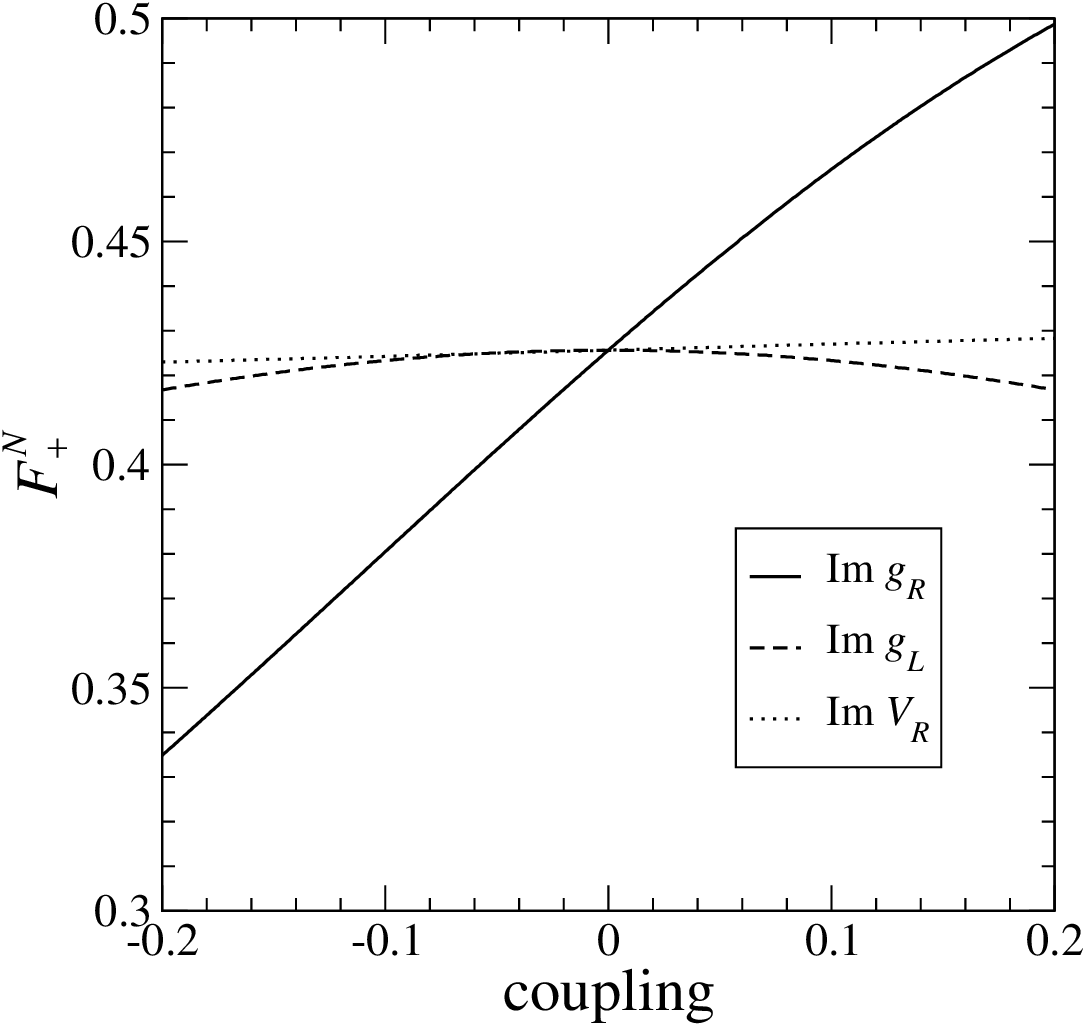,height=4.5cm,clip=}
\end{tabular}
\caption{Dependence of the $W$ polarisation fractions on the imaginary part of $Wtb$ anomalous couplings in Eq.~(\ref{ec:lagr}), taking $\vl = 1$ and real parts vanishing. Up, middle, down: longitudinal, transverse and normal polarisation fractions, respectively.}
\label{fig:FI}
\end{center}
\end{figure}

The transverse and normal polarisation fractions $F_i^T$, $F_i^N$ are defined by normalising to the total width for $t \to Wb$. It is very interesting to observe that they obey a sum rule,
\begin{equation}
\fTz = \fNz = \frac{1}{2} (\fp + \fm) \,,
\label{ec:rel1}
\end{equation}
which can be obtained either from the explicit expressions of the partial widths or by using the relations among polarisation vectors and the fact that $M_{+-}=0$.
Additionally, for a real $Wtb$ vertex,
\begin{equation}
\fNp = \fNm = \frac{1}{2} - \frac{1}{4} (\fp + \fm) \,.
\label{ec:rel2}
\end{equation}
These equations constrain the possible variation of transverse and normal polarisation fractions once that the helicity fractions are measured (see section~\ref{sec:4}).
Their tree-level values in the SM are $\fTm = 0.1718$, $\fTz = 0.1487$, $\fTp = 0.6794$, and $\fNm = 0.4256$, $\fNz = 0.1487$, $\fNp = 0.4256$. For illustration, we show in Figs.~\ref{fig:F} and \ref{fig:FI} the variation of all polarisation fractions for small values of the anomalous couplings, considering only one non-zero anomalous coupling at a time and setting $\vl = 1$ as in the SM. We plot the dependence on the real part of anomalous couplings in Fig.~\ref{fig:F}, whereas the dependence on the imaginary parts is displayed in Fig.~\ref{fig:FI}.%
Comparing both sets of plots we observe that helicity and transverse polarisation fractions are much more sensitive to $\RE \gr$ than to $\IM \gr$, while $\fNpm$ are also very sensitive to $\IM \gr$. Thus, we can anticipate that the eventual measurement of normal $W$ polarisation fractions will significantly improve the constraints on the latter. For a given observable, it is also seen that the dependence on the real and imaginary parts of $\vr$ is similar (but different from one observable to another). The same comment also applies to $\gl$.

\begin{figure}[t]
\begin{center}
\epsfig{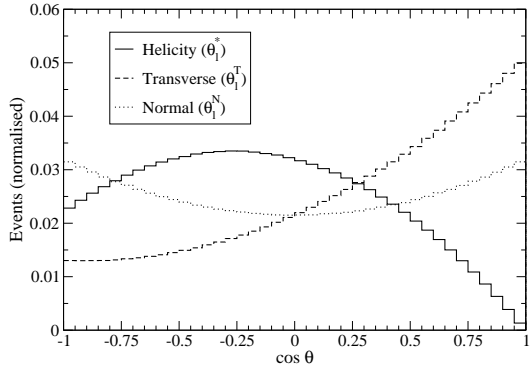}
\caption{Angular distribution of the charged lepton with respect to the three $W$ spin axes: helicity ($\cos \thlw$), transverse ($\cos \thT$) and normal ($\cos \thN$).}
\label{fig:dist}
\end{center}
\end{figure}

As the helicity fractions, the transverse and normal polarisation fractions can be measured in top semileptonic decays. We define the angles $\thT$ ($\thN$) between the charged lepton momentum in the $W$ rest frame and the transverse (normal) directions in the top quark rest frame, given by Eqs.~(\ref{ec:dir}). Then, the charged lepton distribution has the same form as for the angle $\thlw$ in the helicity basis,
\begin{equation}
\frac{1}{\Gamma} \frac{d \Gamma}{d\!\cos \thTN} = \frac{3}{8}
(1 + \cos \thTN)^2 \, \fxp + \frac{3}{8} (1-\cos \thTN)^2 \, \fxm
+ \frac{3}{4} \sin^2 \thTN \, \fxz \,.
\label{ec:distTN}
\end{equation}
The three $\cos \thlw$, $\cos \thT$, $\cos \thN$ distributions are presented in Fig.~\ref{fig:dist} for the SM.
However, in most processes the top quarks are not produced with 100\% polarisation along any axis, but with a certain degree of polarisation
\begin{equation}
P = \frac{N_\uparrow-N_\downarrow}{N_\uparrow+N_\downarrow} \,.
\end{equation}
In this case, the distributions are obtained by substituting in Eq.~(\ref{ec:distTN}) the $W$ polarisation fractions by the ``effective'' quantities
\begin{align}
& \ftxp = \left[ \frac{1+P}{2} \fxp + \frac{1-P}{2} \fxm \right]
\,, \notag
\displaybreak \\
& \ftxm = \left[ \frac{1+P}{2} \fxm + \frac{1-P}{2} \fxp \right] \,, \notag \\ 
& \ftxz = \fxz \,,
\label{ec:distP}
\end{align}
which are the ones actually measured.
Notice that $\fxz$ is unchanged. For an unpolarised top quark ($P = 0$) the resulting distributions are symmetric ($\ftxp = \ftxm$) as one may expect from symmetry arguments. However, the distributions are not isotropic ($\ftxpm \neq \ftxz$) because there is still a privileged direction in space, the $W$ boson momentum.
Experimentally, these distributions can be measured as follows:
\begin{enumerate}
\item In the top quark rest frame, the normal and transverse directions are obtained from Eqs.~(\ref{ec:dir}) using for $\vec s_t$ some spatial direction, preferrably one in which the top quark is produced with a large polarisation ({\em e.g.} the spectator jet momentum in the top rest frame, for $t$-channel single top production~\cite{Mahlon:1999gz}).
\item The momentum of the charged lepton in the $W$ rest frame is obtained performing a boost on its momentum in the top quark rest frame.
\item The angles $\thT$, $\thN$ correspond to the ones between the charged lepton and the two directions previously determined.
\end{enumerate}
We have checked our analytical results for the $W$ polarisation fractions by comparing the predicted distributions with tree-level Monte Carlo calculations in $t$-channel single top production using {\tt Protos}~\cite{AguilarSaavedra:2008gt} and different values of the anomalous couplings, obtaining very good agreement between them.

We conclude this section with a discussion of the corresponding observables for $\bar t \to W^- \bar b$ decays. By explicit calculation it is found that the $W$ polarisation fractions for this decay (denoted with a bar) satisfy
\begin{align}
& \bar F_0 = F_0 \,, \quad \bar F_\pm = F_\mp \,, 
 \notag \\
& \bar F_0^T = F_0^T  \,, \quad \bar F_\pm^T = F_\pm^T \,, \notag \\
& \bar F_0^N = F_0^N  \,, \quad \bar F_\pm^N = F_\pm^N
\label{ec:Fbar}
\end{align}
in full generality, even if the $Wtb$ vertex is CP violating. It is very interesting to observe that CP conservation implies 
\begin{align}
& \bar F_0 = F_0 \,, \quad \bar F_\pm = F_\mp \,, \notag \\
& \bar F_0^T = F_0^T  \,, \quad \bar F_\pm^T = F_\pm^T \,, \notag \\
& \bar F_0^N = F_0^N  \,, \quad \bar F_\pm^N = F_\mp^N \,.
\label{ec:FbarCP}
\end{align}
Then, as expected the longitudinal and transverse polarisation fractions cannot give any information on possible CP-violating effects. On the other hand, for the normal polarisation fractions the simultaneous fulfilment of Eqs.~(\ref{ec:Fbar}) and (\ref{ec:FbarCP}) implies $F_+^N = F_-^N$, as is the case for a CP-conserving $Wtb$ vertex. These relations among polarisation fractions imply that:
\begin{itemize}
\item[(i)] The $\cos \thlw$ distributions are the same for $t$ and $\bar t$ decays because, although the helicity fractions are interchanged, $\bar F_\pm = F_\mp$, the $\cos \theta$ terms in Eqs.~(\ref{ec:dist}) and (\ref{ec:distTN}) also change their sign for $W^-$ decays.
\item[(ii)] For the same reason, the $\cos \thT$ and $\cos \thN$ distributions are also the same provided that the antitop polarisation is the opposite as the one for the top for the axis chosen, $P_{\bar t} = - P_t$.
\end{itemize}

\section{Asymmetries and related observables}
\label{sec:3}

\begin{figure}[t]
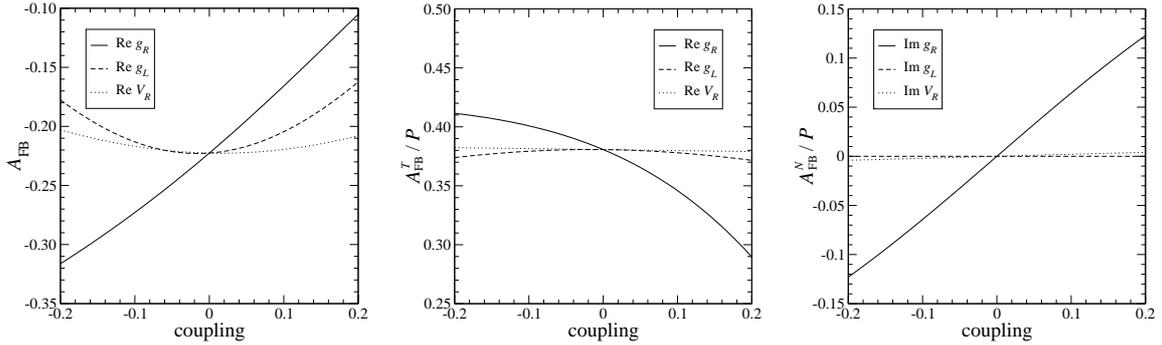

\begin{center}
\begin{tabular}{ccc}
\epsfig{file=Figs/AFB.eps,height=4.5cm,clip=} &
\epsfig{file=Figs/AFBT.eps,height=4.5cm,clip=} &
\epsfig{file=Figs/AFBN.eps,height=4.5cm,clip=}
\end{tabular}
\caption{Left, middle: dependence of the FB asymmetries in the $\cos \thlw$, $\cos \thT$, distributions on the real part of anomalous $Wtb$ couplings in Eq.~(\ref{ec:lagr}), respectively, taking $\vl = 1$ and the rest of anomalous couplings zero.
Right: the same for the $\cos \thN$ distribution and the imaginary parts.
}
\label{fig:AFBN}
\end{center}
\end{figure}

The introduction of the transverse and normal polarisation fractions and the $\cos \thT$, $\cos \thN$ distributions opens the possibility of new angular asymmetries in top quark decays, in complete analogy with the ones obtained for the $\cos \thlw$ distribution~\cite{AguilarSaavedra:2006fy}. One can define asymmetries around any fixed point $z$ in the interval $[-1,1]$,
\begin{equation}
A_z = \frac{N(\cos \theta > z) - N(\cos \theta < z)}{N(\cos \theta > z) +
N(\cos \theta < z)} \,,
\end{equation}
for $\theta = \thlw, \thT, \thN$.
The most obvious choice is $z=0$, giving forward-backward (FB) asymmetries
\begin{align}
& \afb = \frac{3}{4} [\fp - \fm] \,, \notag \\
& \afbTN = \frac{3}{4} [\ftxp - \ftxm] = \frac{3}{4} P [\fxp - \fxm] \,.
\label{ec:afb}
\end{align}
The FB asymmetry in the $\cos \thlw$ distribution $\afb$~\cite{Lampe:1995xb,delAguila:2002nf} does not depend on the top polarisation, while the two 
other ones are proportional to $P$. Their more relevant dependence on anomalous couplings is shown in Fig.~\ref{fig:AFBN}.
The asymmetry $\afbN$, which vanishes for real anomalous couplings (in particular, within the SM), is very sensitive to $\IM \gr$, as it can be seen in the right plot of this figure. For small $\gr$, taking $\vl = 1$, $\vr = \gl = 0$, we obtain
\begin{equation}
\afbN = 0.64 \, P \, \IM \gr \,.
\end{equation}
The numerical coefficient in this asymmetry has also been verified with the Monte Carlo generator {\tt Protos}.
The dependence on $\vr$ is much weaker because it is suppressed by $m_b/m_t$, and the asymmetry does not depend on $\gl$ if the other anomalous couplings vanish.
This asymmetry is the same (up to a minus sign) as the one based on the triple product~\cite{Kane:1991bg}
\begin{equation}
\vec s_t \cdot (\vec p_b \times \vec p_\ell) \,,
\end{equation}
with the $b$ quark and charged lepton momenta taken in the top quark rest frame. Both asymmetries, although sensitive to CP-violating phases in the top decay vertex, are not genuinely CP violating and could be faked by unitarity phases (not considered in our work). The sum of asymmetries for $t$ and $\bar t$ decays,
\begin{equation}
A_\text{FB}^\text{CP} = \afbN(t) + \afbN(\bar t)
\end{equation}
is unambiguously CP violating.

It is worthwhile to remark here that $\afbN$ can be relatively large because it
directly probes the imaginary parts of the off-diagonal density matrix elements for a polarised top quark decay, namely $D_1$ in Eqs.~(\ref{ec:AtoF}). Therefore, it is expected to be much larger than CP-violating
asymmetries based on triple-product spin correlations in $t \bar t$ production~\cite{Brandenburg:1992be,Bernreuther:1993hq,Antipin:2008zx,Gupta:2009wu,Gupta:2009eq}.
Using {\tt Protos} for $t \bar t$ generation with anomalous $Wtb$ couplings,\footnote{This generator has been thoroughly tested, validated and is used for official production of $t \bar t$ samples with anomalous $Wtb$ couplings in ATLAS.}
we actually find CP asymmetries numerically much smaller (up to a factor of 35) than the ones obtained in Ref.~\cite{Gupta:2009wu}. For example,
\begin{align}
& \tilde A_1 = (0.0886 \pm 0.0015) \, \IM \gr
= (-0.0407 \pm 0.0007) f_t \sin \phi_f  \,, \notag \\
& \tilde A_2 = (0.0191 \pm 0.0015) \, \IM \gr 
= (-0.0087 \pm 0.0007) f_t \sin \phi_f \,, \notag \\
& \tilde A_3 = (0.0328 \pm 0.0015) \, \IM \gr
= (-0.0150 \pm 0.0007) f_t \sin \phi_f \,,
\end{align}
where $\tilde A_{1-3}$ have been defined in Ref.~\cite{Gupta:2009wu} and $\IM \gr = - M_W/m_t f_t \sin \phi_f$ in their notation. The uncertainties quoted come from the Monte Carlo statistics. The numerical results we obtain for $\tilde A_{1-3}$ seem consistent with the expectation that spin correlation asymmetries, in particular the CP-violating ones, are suppressed by the spin correlation between the top and antitop, among other factors. (See Eqs.~(\ref{ec:Ath}) in section~\ref{sec:7} for CP-conserving correlations.)

Other convenient choices for asymmetries in the $\cos \theta$ distributions are $z = \mp (2^{2/3}-1)$. Defining
$\beta = 2^{1/3}-1$, we have
\begin{align}
& z = -(2^{2/3}-1) & \rightarrow \quad & \Ap = 3 \beta [\fz+(1+\beta) \fp] \,,
\notag \\[1mm]
& & & \Apx = 3 \beta [\fxz+(1+\beta) \ftxp] \,, \notag \\[1mm]
& z = (2^{2/3}-1) & \rightarrow \quad & \Am = -3 \beta [\fz+(1+\beta) \fm] \,, \notag \\[1mm]
& & & \Amx = -3 \beta [\fxz+(1+\beta) \ftxm] \,.
\label{ec:apm}
\end{align}
The resulting asymmetries only depend on two ``effective'' polarisation fractions. Conversely, the latter quantities can also be determined from asymmetries, for example
\begin{eqnarray}
\ftxp & = & \frac{1}{1-\beta} + \frac{\Amx - \beta \Apx}{3 \beta(1-\beta^2)} \,,
 \notag \\
\ftxm & = & \frac{1}{1-\beta} - \frac{\Apx - \beta \Amx}{3 \beta(1-\beta^2)} \,,
 \notag \\
\fxz & = & - \frac{1+\beta}{1-\beta} + \frac{\Apx - \Amx}{3 \beta (1-\beta)} \,.
\label{ec:inv}
\end{eqnarray}
The angular asymmetries $\afbTN$, $\Apx$, $\Amx$ do not provide any further information than the polarisation fractions $\ftxp$, $\ftxm$, $\fxz$ do. Still, their measurement may be more convenient from the experimental point of view, especially with low statistics, since it does not require fitting the $\cos \thTN$ distributions. Moreover, systematic uncertainties on the asymmetries may be smaller than on the polarisation fractions,\footnote{For a detailed comparison of systematic uncertainties on helicity fractions, their ratios and angular asymmetries
see Ref.~\cite{AguilarSaavedra:2007rs}.}
so that the constraints placed on anomalous $Wtb$ couplings may be stronger. A detailed evaluation of systematic uncertainties for these measurements is compulsory before drawing any conclusion in this respect.

\section{Indirect constraints on polarisation fractions}
\label{sec:4}

As we have remarked, the sum rule in Eq.~(\ref{ec:rel1}) implies that the measurement of helicity fractions in top quark decays automatically fixes the $\fTz$, $\fNz$ components of the transverse and normal $W$ polarisation. Still, the other four components are undetermined in principle. We have investigated their possible range of variation given the present Tevatron measurements~\cite{Aaltonen:2010ha} and the future expectations for LHC with 10 fb$^{-1}$ at a centre of mass (CM) energy of 14 TeV~\cite{AguilarSaavedra:2007rs}. We take 
\begin{align}
& \fz = 0.88 \pm 0.125 \,, \quad \fp = -0.15 \pm 0.0921 \,, \quad \text{corr} = -0.59 
&& \text{(Tevatron)} \,, \notag \\
& \fz = 0.700 \pm  0.0192 \,, \quad \fp = 0.0006 \pm 0.00216
&& \text{(LHC)} \,.
\label{ec:data1}
\end{align}
For the forthcoming LHC measurements the correlation has not yet been estimated and is therefore ignored. We also use single top cross section measurements, which constrain the anomalous couplings in Eq.~(\ref{ec:lagr}) and then, indirectly, the $W$ polarisation fractions.
For Tevatron we use the combined $s+t$-channel measurement~\cite{Aaltonen:2009jj} which has a better precision than the separate ones for the $s$- and $t$-channels. For LHC we restrict ourselves to $tW$ production, which does not receive contributions from other types of new physics, for example four-fermion operators, and probes the $Wtb$ vertex in a model-independent fashion. (Limits on anomalous couplings from $s$- and $t$-channel measurements could be relaxed by the introduction of four-fermion operators also contributing to the production amplitudes.) We take the values
\begin{align}
& \sigma_t + \sigma_s = 2.3^{+0.6}_{-0.5} ~\text{pb}
&& \text{(Tevatron)} \,, \notag \\
& \sigma_{tW} = 66 \pm 13 ~\text{pb}
&& \text{(LHC)} \,.
\label{ec:data2}
\end{align}
The fits are performed using {\tt TopFit\,2} letting the four couplings in the $Wtb$ Lagrangian arbitrary.\footnote{ Our extraction of limits from cross sections does not take into account the variation of the event selection efficiency when anomalous couplings are introduced, which requires a detailed simulation. Nevertheless, for the results presented here this effect is expected to have little relevance.}
We generate random points in the $(\vl,\vr,\gl,\gr)$ parameter space with a flat probability distribution and use the acceptance-rejection method to obtain a sample distributed according to the combined $\chi^2$ of the observables considered. The limits presented are $1\sigma$ regions with a boundary of constant $\chi^2$ containing 68.26\% of the points accepted. A more detailed description of the method used can be found in Ref.~\cite{AguilarSaavedra:2006fy}.
We show in turn the results for the CP-conserving case (all anomalous couplings real) and for a general complex $Wtb$ vertex. This distinction is partially motivated by the fact that the imaginary parts of anomalous $Wtb$ couplings generated at one loop level 
in popular SM extensions are rather small~\cite{Bernreuther:2008us}. Besides, we note that there are observables such as the ratio $\rp = \fp/\fz$ and the asymmetry $\Ap$ (see the previous section) which are more constraining than helicity fractions themselves, but the limits on $\fTpm$, $\fNpm$ obtained using them are practically the same, and for simplicity we use the expected helicity fraction measurements.

\begin{figure}[h]
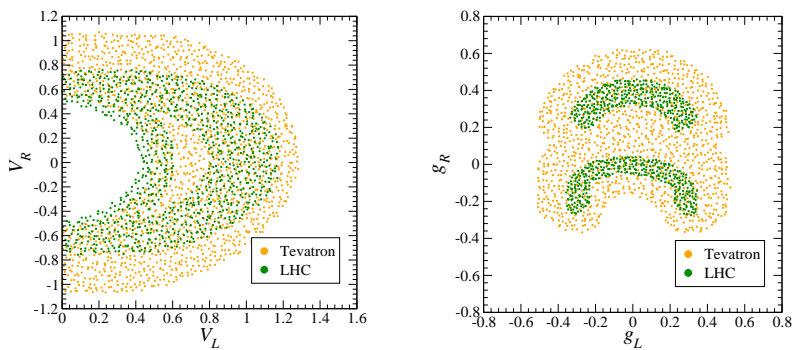

\begin{center}
\begin{tabular}{ccc}
\epsfig{file=Figs/lim-Wh+st-R-vlvr.eps,height=4.5cm,clip=} & \quad &
\epsfig{file=Figs/lim-Wh+st-R-glgr.eps,height=4.5cm,clip=}
\end{tabular}
\caption{Combined limits on anomalous couplings, assumed real, obtained from the (expected) measurements in Eqs.~(\ref{ec:data1}) and (\ref{ec:data2}).}
\label{fig:limR}
\end{center}
\end{figure}

The limits on real anomalous couplings are shown in Fig.~\ref{fig:limR}.
The two plots are projections of the four-dimensional $1\sigma$ region obtained, allowing for all cancellations among the different terms. In particular, the upper green (dark gray) area in the right plot corresponds to a large cancellation between the linear $\vl \gr^*$ terms, which are not suppressed by the $b$ quark mass, and the quadratic ones $|\gr|^2$. This cancellation is also seen in the $(\RE \gr,\IM \gr$) plane, for the general complex case discussed below. We point out that, despite the good precision of helicity fraction measurements, the limits obtained here are rather loose due to cancellations among different contributions involving more than one non-zero anomalous coupling and/or
$\vl < 1$.%
\begin{figure}[t]
\begin{center}
\epsfig{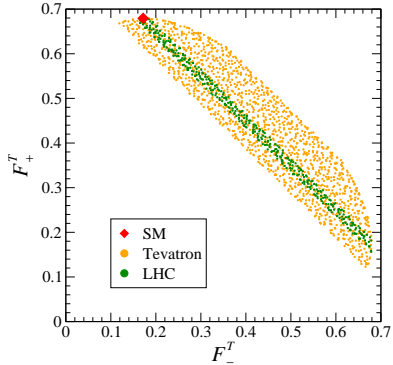}
\caption{Variation of the transverse polarisation fractions $\fTm$, $\fTp$ for $Wtb$ couplings in the $1\sigma$ regions of Fig.~\ref{fig:limR}, for real anomalous couplings.}
\label{fig:Wrng-R}
\end{center}
\end{figure}
The variation of $\fTpm$ for $Wtb$ couplings inside these regions is shown in Fig.~\ref{fig:Wrng-R}. We also mark the value corresponding to the SM prediction. Notice that when anomalous couplings are present $\fTp$ takes values smaller than the SM one, in agreement with Fig.~\ref{fig:F}. This plot demonstrates that, given the present (and expected) constraints on helicity fractions, there is still large room for departures from the SM prediction for $\fTp$, $\fTm$. Hence, their measurement 
is necessary and will provide useful constraints on the $Wtb$ vertex.
For real anomalous couplings, the normal polarisation fractions are fixed by the sum rule in Eq.~(\ref{ec:rel2}) once that helicity fractions are measured, and the corresponding plot is not shown. 

\begin{figure}[h]
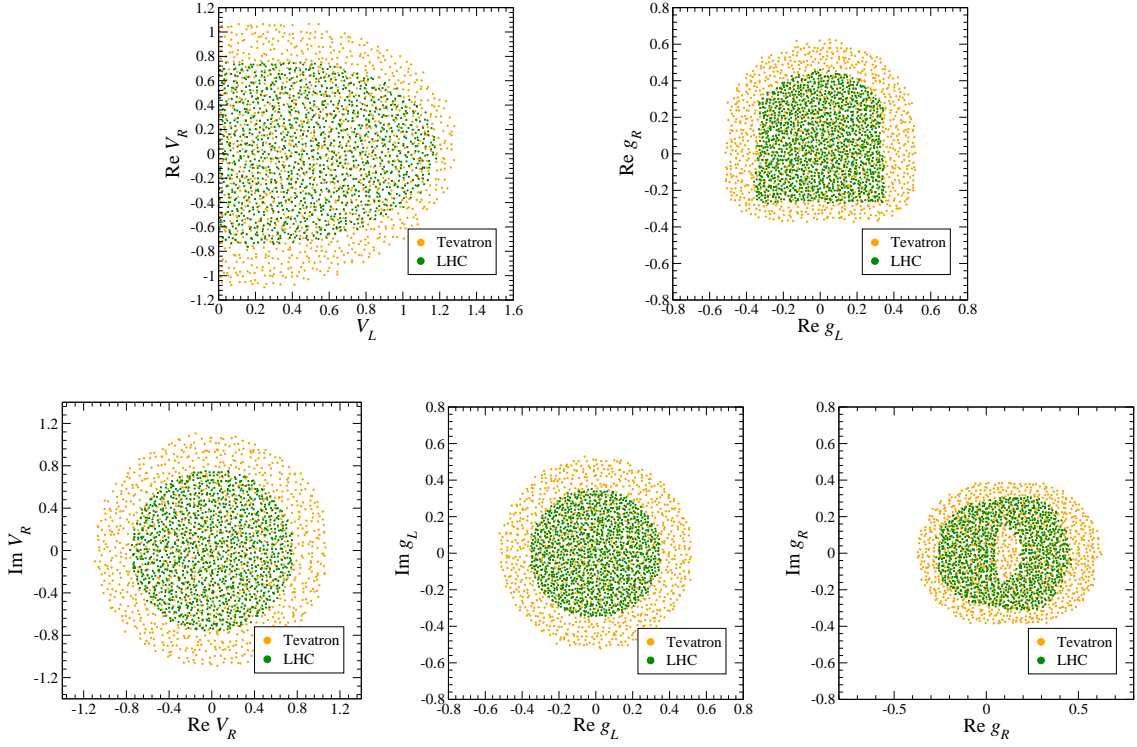

\begin{center}
\begin{tabular}{ccc}
\multicolumn{3}{c}{\begin{tabular}{ccc}
\epsfig{file=Figs/lim-Wh+st-C-vlvr.eps,height=4.5cm,clip=} & \quad \quad &
\epsfig{file=Figs/lim-Wh+st-C-glgr.eps,height=4.5cm,clip=}
\end{tabular} } \\ \\
\epsfig{file=Figs/lim-Wh+st-C-vrwr.eps,height=4.5cm,clip=} & 
\epsfig{file=Figs/lim-Wh+st-C-glhl.eps,height=4.5cm,clip=} &
\epsfig{file=Figs/lim-Wh+st-C-grhr.eps,height=4.5cm,clip=}
\end{tabular}
\caption{Combined limits on anomalous couplings obtained from the (expected) measurements in Eqs.~(\ref{ec:data1}) and (\ref{ec:data2}), for the general complex case.}
\label{fig:limC}
\end{center}
\end{figure}

The $1\sigma$ limits on anomalous couplings for the general case are presented in Fig.~\ref{fig:limC}. On the upper row we show the limits on $\vl$ (taken real and positive by definition) and the real parts of $\vr$, $\gl$ and $\gr$. These plots correspond to the ones shown in Fig.~\ref{fig:limR} but the allowed regions are larger, because with three more free parameters in the fits there is more room for cancellations among different contributions. In the lower row we show the limits on the real and imaginary parts of the anomalous couplings $\vr$, $\gl$ and $\gr$.
For the first two, helicity fractions and single top cross sections basically set limits on $|\vr|^2$ and $|\gl|^2$, respectively.\footnote{This fact does not contradict our previous claim that linear terms in $\vr$, $\gl$ proportional to the $b$ quark mass are important, because here the limits are rather loose due to the few number of observables included and the possibility of cancellations. Indeed, the important effect of the $b$ quark mass can be clearly appreciated in the results presented in section~\ref{sec:8}.} The limits on $\gr$, for a fixed $\vl$, have a ring shape in the $(\RE \gr,\IM \gr$) plane. The resulting regions in Fig.~\ref{fig:limC} (down, right) are the superposition of several such rings of different centres and radii.%
\begin{figure}[t]
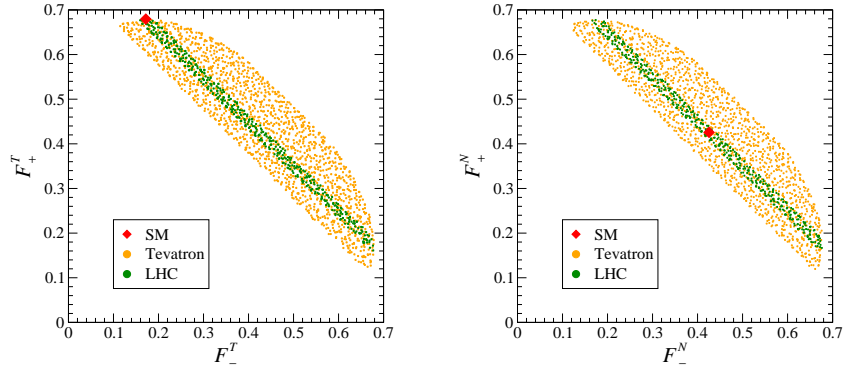

\begin{center}
\begin{tabular}{ccc}
\epsfig{file=Figs/WTrng-C.eps,height=4.8cm,clip=} & \quad &
\epsfig{file=Figs/WNrng-C.eps,height=4.8cm,clip=}
\end{tabular}
\caption{Variation of the transverse polarisation fractions $\fTm$, $\fTp$ for $Wtb$ couplings in the $1\sigma$ regions of Fig.~\ref{fig:limC}, for complex anomalous couplings.}
\label{fig:Wrng-C}
\end{center}
\end{figure}
The variation of $\fTpm$, $\fNpm$ for $Wtb$ couplings inside these regions is shown in Fig.~\ref{fig:Wrng-C}. We also mark the values corresponding to the SM prediction. For $\fTpm$ the allowed range is very large, practically the same as in the real case. For $\fNpm$, we also observe that there is ample room for departures from the SM equality $\fNp = \fNm$. Therefore, their determination is quite interesting in order to explore new physics contributing to the $Wtb$ vertex, in particular if the anomalous couplings have complex phases.

\section{$W$ polarisation and angular distributions in the top quark rest frame}
\label{sec:5}

The presence of anomalous $Wtb$ couplings influences the angular distribution of the $W$ boson produced in the decay $t \to Wb$, in addition to its polarisation. Indirectly, the angular distribution in the top quark rest frame of the $W$ decay products is affected by both. Then, it is pertinent to ask ourselves about the relation between the measurement of transverse and normal $W$ polarisations and the distributions of top quark decay products.

\begin{figure}[h]
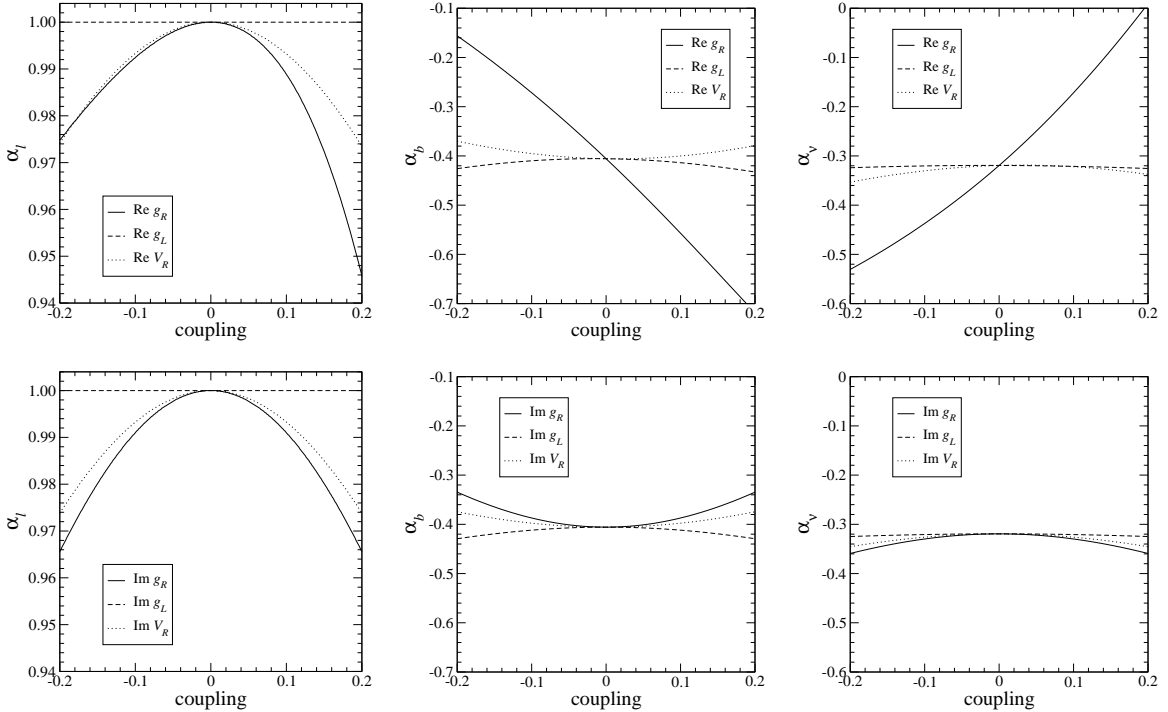

\begin{center}
\begin{tabular}{ccc}
\epsfig{file=Figs/al.eps,height=4.5cm,clip=} &
\epsfig{file=Figs/ab.eps,height=4.5cm,clip=} &
\epsfig{file=Figs/an.eps,height=4.5cm,clip=} \\[2mm]
\epsfig{file=Figs/ali.eps,height=4.5cm,clip=} &
\epsfig{file=Figs/abi.eps,height=4.5cm,clip=} &
\epsfig{file=Figs/ani.eps,height=4.5cm,clip=}
\end{tabular}
\caption{Dependence of the spin analysing powers on the real (up) and imaginary (down) part of $Wtb$ anomalous couplings in Eq.~(\ref{ec:lagr}), taking $\vl = 1$.}
\label{fig:a}
\end{center}
\end{figure}

For the decay $t \to W^+ b \to \ell^+ \nu b,q \bar q' b$, the angular
distribution of any decay product $X=\ell^+,\nu,q,\bar q',W^+,b$ (which are
called ``spin analysers'') in the top quark rest frame is given by
\begin{equation}
\frac{1}{\Gamma} \frac{d\Gamma}{d\!\cos \theta_X} = \frac{1}{2} (1+\alpha_X \cos
\theta_X )
\label{ec:tdist}
\end{equation}
with $\theta_X$ the angle between the three-momentum of $X$ in the $t$
rest frame and the top spin direction.
The constants $\alpha_X$ are called ``spin analysing power'' of $X$ and can range
between $-1$ and $1$. In the SM $\alpha_{\ell^+} = \alpha_{\bar q'} = 1$, $\alpha_\nu = \alpha_q = -0.32$ and $\alpha_b = - \alpha_{W^+} = - 0.41$ at the tree level~\cite{Jezabek:1994qs} ($q$ and $q'$ are the up- and down-type quarks, respectively, resulting from the $W$ decay). One-loop corrections slightly modify these values to
$\alpha_{\ell^+} = 0.998$, $\alpha_{\bar q'} = 0.93$,
$\alpha_\nu = -0.33$, $\alpha_q = -0.31$, $\alpha_b = - \alpha_{W^+} = - 0.39$
\cite{Czarnecki:1994bn,Brandenburg:2002xr,Bernreuther:2004jv}.
We have calculated the spin analysing power constants
for the general (complex) $Wtb$ vertex in Eq.~(\ref{ec:lagr}), keeping $m_b$ non-zero and quadratic terms in the couplings, generalising previous results in the literature~\cite{Grzadkowski:1999iq,Godbole:2006tq,AguilarSaavedra:2006fy}.
The spin analysing power constants can be written as $\alpha_X = a_X/a_0$, with
\begin{align}
a_0 & = 
 \left[ |\vl|^2 + |\vr|^2 \right] \left(1 + x_W^2 -2 x_W^4 \right) 
 + 2 \left[ |\gl|^2 + |\gr|^2 \right] 
 \left(1 - \frac{x_W^2}{2}  - \frac{x_W^4}{2}  \right) \notag \\
& - 12 x_W^2 x_b \, \RE \left[ \vl \vr^* + \gl \gr^* \right]
- 6 x_W \RE \left[\vl \gr^* + \vr \gl^* \right]
\left(1 - x_W^2 \right) \notag \\
& + 6 x_W x_b \, \RE \left[\vl \gl^* + \vr \gr^* \right] \,, \notag \\
%%%%%%%%%%%%%%
a_b & = -2 \frac{\qb}{m_t} \left\{ \left[ |\vl|^2 - |\vr|^2 \right] 
\left( 1-2 x_W^2 \right)
+ 2 \left[ |\gl|^2 - |\gr|^2 \right] \left( 1 - \frac{x_W^2}{2} \right)
\right. \notag \\
& \left. +2 x_W \, \RE \left[ \vl \gr^* - \vr \gl^* \right]
+ 6 x_W x_b \, \RE \left[ \vl \gl^* - \vr \gr^* \right] \right\}   \,, \notag
 \\
%%%%%%%%%%%%%%%
a_{\ell^+} & =
   \left[ |\vl|^2 - |\vr|^2 \right] \left(1 + x_W^2 -2 x_W^4 \right)
   + 2 \left[ |\gl|^2 - |\gr|^2 \right]
\left(1 - \frac{x_W^2}{2} - \frac{x_W^4}{2} \right)  \notag \\
& - 12 x_W^2 x_b \, \RE \left[ \vl \vr^* + \gl \gr^* \right] 
- 6 x_W \, \RE \left[\vl \gr^* + \vr \gl^* \right]
\left(1 - x_W^2 \right) \notag \\
& + 6 x_W x_b \, \RE \left[\vl \gl^* - \vr \gr^* \right]
\left(1 +x_W^2 \right) + 12 x_W^2 \left[ |\vr|^2 - |\gr|^2 \right] \notag
% \displaybreak
 \\
& + 6 \frac{M_W}{\qb} x_W \log \frac{E_W+\qb}{E_W-\qb}
\left[ |\gr|^2 -x_W^2 |\vr|^2 +2 x_W x_b \, \RE \vr \gr^*
\right] \,, 
 \displaybreak \notag \\
%%%%%%%%%%%%%%
a_{\nu} & = \left[ |\vl|^2 - |\vr|^2 \right] \left(1 + x_W^2 -2 x_W^4 \right)
+ 2 \left[ |\gl|^2 - |\gr|^2 \right]
\left(1 - \frac{x_W^2}{2} - \frac{x_W^4}{2} \right) \notag \\
& + 12 x_W^2 x_b \, \RE \left[ \vl \vr^* + \gl \gr^* \right] 
+ 6 x_W \, \RE \left[\vl \gr^* + \vr \gl^* \right]
\left(1 - x_W^2 \right) \notag \\
& + 6 x_W x_b \, \RE \left[\vl \gl^* - \vr \gr^* \right]
\left(1 +x_W^2 \right) - 12 x_W^2 \left[ |\vl|^2 - |\gl|^2 \right] \notag \\
& - 6 \frac{M_W}{\qb} x_W \log \frac{E_W+\qb}{E_W-\qb}
\left[ |\gl|^2 - x_W^2 |\vl|^2 +2 x_W x_b \, \RE \vl \gl^*
\right] \,,
%%%%%%%%%%%%%%
\label{ec:kappas}
\end{align}
with $E_W^2 = M_W^2 + \qb^2$.
In the above expressions we have omitted for brevity terms of order $x_b^2$ and higher, which are kept in our numerical code anyway.
For the rest of top quark decay products we have $a_{\bar q'} = a_ {\ell^+}$, $a_q = a_\nu$ and $a_{W^+} = - a_b$.
The angular distributions for the decay of a top antiquark are the same, with $\alpha_{\bar X} = - \alpha_X$ even in the CP-violating case, as obtained by an explicit calculation.
We point out that imaginary parts of coupling products do not directly enter these expressions (only through the moduli squared). Although they appear in the matrix element squared involved in the evaluation of $a_\ell$ and $a_\nu$ they cancel when integrated over the azimuthal angle $\psi_X$ of the the spin analyser momentum with respect to the top spin. The dependence of spin analysing powers on the real and imaginary parts of anomalous couplings is shown in Fig.~\ref{fig:a}.
We consider only one anomalous coupling non-zero at a time, and show separately the dependence on the real and imaginary parts.

For partially polarised top quark decays the distribution in Eq.~(\ref{ec:tdist}) is modified to
\begin{equation}
\frac{1}{\Gamma} \frac{d\Gamma}{d\!\cos \theta_X} = \frac{1}{2} (1+ P \alpha_X \cos
\theta_X ) \,,
\label{ec:tdistP}
\end{equation}
so that the quantities actually measured in the distributions are the products $P \alpha_X$. For example, the FB asymmetries
\begin{equation}
A_X = \frac{N(\cos \theta_X > 0) - N(\cos \theta_X < 0)}{N(\cos \theta_X > 0) +
N(\cos \theta_X < 0)}
\label{ec:Aspin}
\end{equation}
are $A_X = P \alpha_X / 2$. A first estimate of the precision in the measurement of these asymmetries in $t$-channel single top production, including systematic uncertainties, has been given in Ref.~\cite{AguilarSaavedra:2008zz}. However, due to the smallness of the available simulated samples the uncertainties seem to be overestimated. We will then assume an improvement by a factor of two in the systematic uncertainties, and take the statistical ones as $\sqrt{1-A_X^2}/\sqrt N$, with $N$ the number of signal events. The resulting sensitivities are:
\begin{align}
& \Delta A_\ell  = 0.012~\text{(stat)} \oplus 0.016~\text{(sys)} \,, \notag \\
& \Delta A_b = 0.013~\text{(stat)} \oplus 0.011~\text{(sys)} \,, \notag \\
& \Delta A_\nu = 0.013~\text{(stat)} \oplus 0.017~\text{(sys)}  \,.
\label{ec:Ameas}
\end{align}
They correspond to 4.6\%, 8.9\% and 14.6\% relative precisions in the asymmetry measurements, which do not seem too optimistic.
In the rest of this section we investigate the relation between the measurement of transverse polarisation fractions and spin analysing power constants.
As we have done before, we distinguish the cases of real and complex anomalous couplings.

\begin{figure}[htb]
\begin{center}
\epsfig{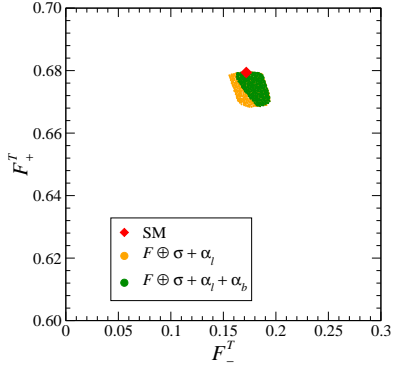}
\caption{Variation of the transverse polarisation fractions $\fTm$, $\fTp$ for $Wtb$ couplings in the $1\sigma$ regions of Fig.~\ref{fig:limR} (real $Wtb$ vertex) corresponding to LHC limits, also including constraints on $\alpha_b$ and/or $\alpha_\ell$ (see the text).}
\label{fig:Wrng-Ra}
\end{center}
\end{figure}

For a real $Wtb$ vertex, a precise measurement of spin analysing powers (together with helicity fractions and the $tW$ cross section) significantly shrinks the allowed region for $\fTm$, $\fTp$. We have required, for the points in the $1\sigma$ region in Fig.~\ref{fig:limR} corresponding to LHC limits, that: (i) $\alpha_\ell$ is between 4.6\% of its SM value; (ii) the same, plus $\alpha_b$ within 8.9\% of its SM value. (These precisions correspond to the ones of the respective spin asymmetries.) The results are shown in Fig.~\ref{fig:Wrng-Ra}. An additional measurement of $\alpha_\nu$ with the expected precision would have negligible impact on the variation of $\fTm$, $\fTp$. Although the possible variation of $\fTpm$ is much more constrained in this case we
see that, given the expected uncertainties, it is likely that this measurement will be complementary to asymmetries in top quark rest frame.

\begin{figure}[htb]
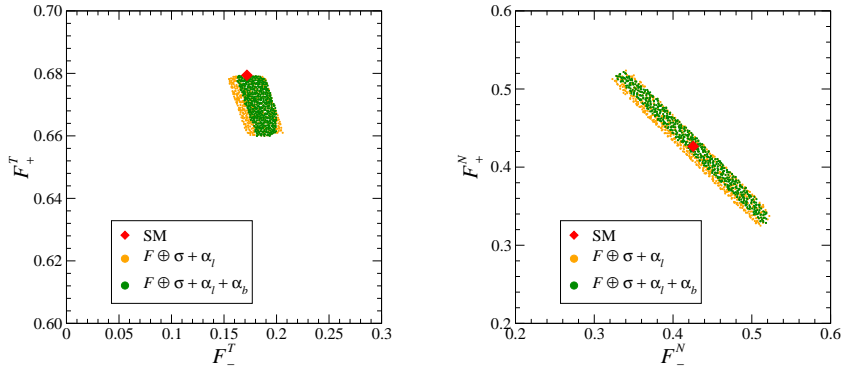

\begin{center}
\begin{tabular}{ccc}
\epsfig{file=Figs/WTrng-Ca.eps,height=4.8cm,clip=} & \quad &
\epsfig{file=Figs/WNrng-Ca.eps,height=4.8cm,clip=}
\end{tabular}
\caption{Variation of the transverse polarisation fractions $\fTm$, $\fTp$ for $Wtb$ couplings in the $1\sigma$ regions of Fig.~\ref{fig:limC} (complex $Wtb$ vertex) corresponding to LHC limits, also including constraints on $\alpha_b$ and/or $\alpha_\ell$ (see the text).}
\label{fig:Wrng-Ca}
\end{center}
\end{figure}

For a general complex $Wtb$ vertex the results for $\fTpm$ are slightly different, as it is shown in Fig.~\ref{fig:Wrng-Ca} (left). After imposing constraints on $\alpha_\ell$ and $\alpha_b$ (an additional requirement on $\alpha_\nu$ does not make any difference) the range of variation of $\fTpm$ is roughly two times larger. More importantly, the normal polarisation fractions can have significant departures from their SM prediction (right panel). This fact justifies the necessity of their future measurement at LHC, for example in $t$-channel single top production. 

\section{Model-independent fit of the $Wtb$ vertex}
\label{sec:6}

In section~\ref{sec:4} we have estimated the future LHC limits on $Wtb$ couplings only using helicity fraction measurements and the $tW$ cross section. As we have remarked, these limits are somewhat loose (despite the good precision expected for helicity fractions) due to the possibility of cancellations among different contributions when more than one anomalous coupling is non-zero. Cancellations can be reduced, and limits can be greatly improved, by including in the fits observables related to the top quark polarisation, which are sensitive to additional entries in the spin density matrix.
These observables, which are expected to be measured with good accuracy in $t$-channel single top production, include the spin asymmetries in the top quark rest frame, defined in Eqs.~(\ref{ec:Aspin}), the transverse/normal polarisation fractions and related asymmetries. 
All of them depend on the (a priori unknown) $t$-channel single top quark polarisation, which is taken as a free parameter and is obtained from the fit.\footnote{One could still argue that the single top polarisation is calculable in terms of the $Wtb$ couplings. However, this assumes that no other new physics, {\em e.g.} four-fermion operators, contributes to $t$-channel production, and turns the results model-dependent.}

The main purpose of the fits performed in this section is, rather than providing very precise estimates of the LHC sensitivity to $Wtb$ anomalous couplings, to show that a simultaneous measurement of all $Wtb$ couplings and the single top polarisation is feasible, and that results are greatly improved by using normal $W$ polarisation observables. For our fits we use the helicity fraction measurements, for consistency with sections~\ref{sec:4} and \ref{sec:5}, and the $tW$ cross section.
(We note, however, that using observables such as the ratio $\rp = \fp/\fz$ and the asymmetry $\Ap$ the limits on anomalous couplings might be improved up to 30\%~\cite{AguilarSaavedra:2007rs}, but this is not crucial for our discussion.)
In addition, we include the spin asymmetries $A_\ell$, $A_b$ involving the charged lepton and $b$ quark distribution in the top quark rest frame, with the sensitivities given in Eqs.~(\ref{ec:Ameas}). For the transverse and normal polarisation we use the FB asymmetries $\afbTN$, whose measurement is expected to have similar precision as the usual FB asymmetry in the $\cos \thlw$ distribution. We will then take~\cite{AguilarSaavedra:2008zz}
\begin{align}
& \Delta \afbTN  = 0.013~\text{(stat)} \oplus 0.018~\text{(sys)} \,.
\label{ec:Ameas2}
\end{align}
We point out that the measurement of $\fTpm$, $\fNpm$ themselves (whose accuracy is difficult to estimate with present simulations) might be more constraining and yield better bounds on the $Wtb$ vertex. The ``experimental'' values which are used for the fits correspond to the SM prediction assuming a top polarisation $P = 0.9$, close to (but conservatively smaller than) the one predicted for $t$-channel single top production~\cite{Mahlon:1999gz}.

\begin{figure}[htb]
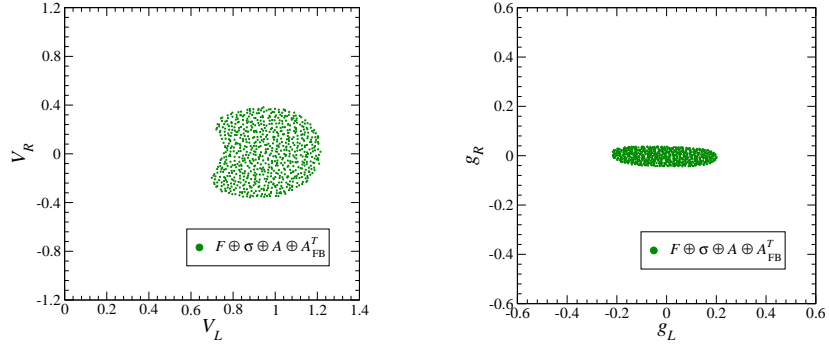

\begin{center}
\begin{tabular}{ccc}
\epsfig{file=Figs/fit-R-vlvr.eps,height=4.5cm,clip=} & \quad \quad &
\epsfig{file=Figs/fit-R-glgr.eps,height=4.5cm,clip=}
\end{tabular}
\caption{Combined limits on anomalous couplings obtained from expected measurements at LHC, assuming a real $Wtb$ vertex.}
\label{fig:fitR}
\end{center}
\end{figure}

\begin{figure}[htb]
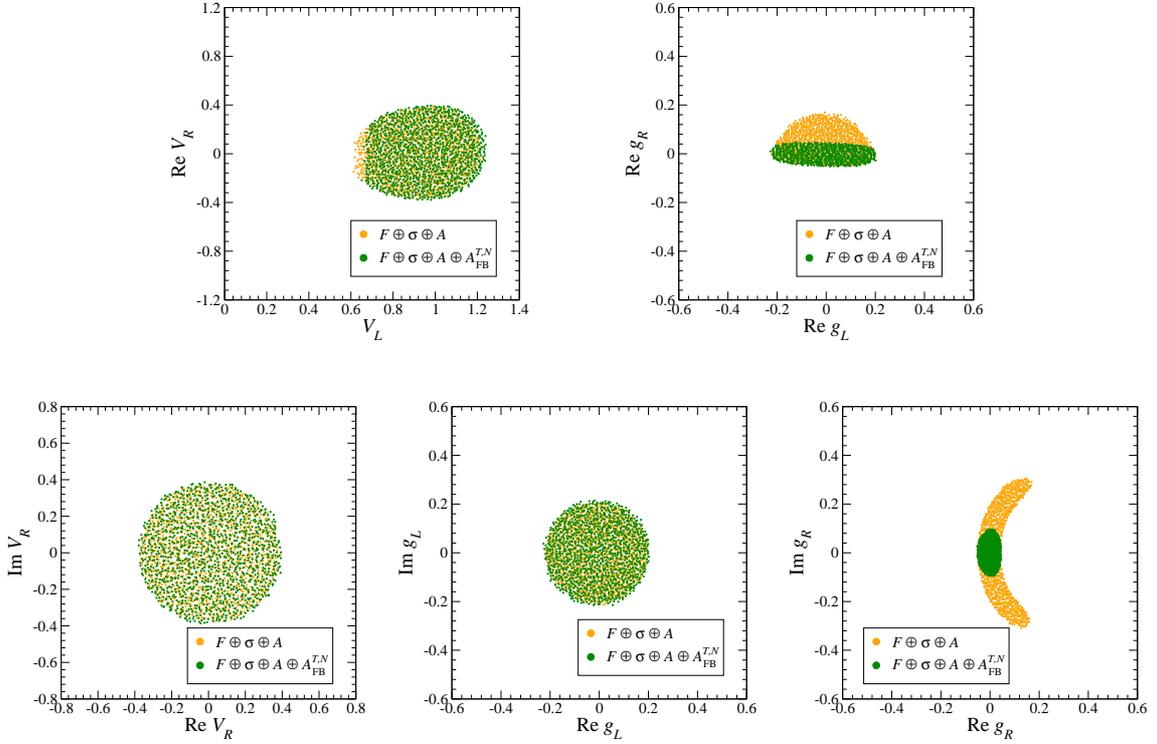

\begin{center}
\begin{tabular}{ccc}
\multicolumn{3}{c}{\begin{tabular}{ccc}
\epsfig{file=Figs/fit-C-vlvr.eps,height=4.5cm,clip=} & \quad \quad &
\epsfig{file=Figs/fit-C-glgr.eps,height=4.5cm,clip=}
\end{tabular} } \\ \\
\epsfig{file=Figs/fit-C-vrwr.eps,height=4.5cm,clip=} &
\epsfig{file=Figs/fit-C-glhl.eps,height=4.5cm,clip=} &
\epsfig{file=Figs/fit-C-grhr.eps,height=4.5cm,clip=}
\end{tabular}
\caption{Combined limits on anomalous couplings obtained from expected measurements at LHC, for a general complex $Wtb$ vertex.}
\label{fig:fitC}
\end{center}
\end{figure}

The results of the fit assuming a real $Wtb$ vertex are shown in Fig.~\ref{fig:fitR}. We do not include the asymmetry $\afbN$, which identically vanishes in this case. The results for the general complex $Wtb$ vertex are presented in Fig.~\ref{fig:fitC}.%
The more remarkable features of these plots are:
\begin{itemize}
\item[(i)] The limits on $\gr$ are quite precise, below the 5\% level, and take advantage of the good sensitivity of $W$ polarisation fractions to this coupling, due to the
large interferences $\RE \vl \gr^*$, $\IM \vl \gr^*$ with the SM coupling. The improvement with respect to the results in section~\ref{sec:4} is quite remarkable.
\item[(ii)] In the complex case, the inclusion of the asymmetry $\afbN$, very sensitive to $\IM \vl \gr^*$, leads to a significant improvement of the limits on $\IM \gr$ and, indirectly, on $\RE \gr$, with respect to the limits without these asymmetries shown in orange (light gray).
\item[(iii)] In both cases the model-independent determination of the SM coupling $\vl$ from the combined fit has a relatively good precision, although worse than the model-dependent one obtained from $tW$ production alone assuming that no anomalous $Wtb$ couplings exist.
\end{itemize}
The single top polarisation is also obtained from the fit, and ranges in the interval $[0.83,1]$ in both cases. This polarisation tests the presence of new physics in single top production, such as new gauge bosons, top flavour-changing neutral couplings or four-fermion operators.

\section{Implications for spin correlations in $t \bar t$ production}
\label{sec:7}

We address here the implications of our results for spin correlations in $t \bar t$ production at LHC. As it is well known, the spin correlation between the top and antitop can be modified by the presence of new production mechanisms~\cite{Dicus:1994bm,Bernreuther:1997gs,Choudhury:2007ux} and thus it probes new physics in $t \bar t$ production. Still, this correlation is observable through the distributions of $t$, $\bar t$ decay products which, as we have discussed in the preceding sections, are sensitive to new physics in the decay vertex, {\em i.e.} anomalous $Wtb$ couplings.

In $t \bar t$ production top quarks are produced unpolarised at the tree
level in QCD interactions, and with a very small $O(10^{-2})$ transverse polarisation at one loop. However, the $t$
and $\bar t$ spins are correlated, allowing for the construction of
angular asymmetries at the percent level.
Working in the helicity basis and neglecting small spin interference
effects, so that the cross section factorises into production times decay
factors, the double angular distribution of the decay products $X$ (from $t$)
and $\bar X'$ (from $\bar t$) can be written as~\cite{Mahlon:1995zn,Stelzer:1995gc}
\begin{equation}
\frac{1}{\sigma} \frac{d\sigma}{d\!\cos \theta_X \, d\!\cos \theta_{\bar X'}} =
\frac{1}{4} (1+C \, \alpha_X \alpha_{\bar X'} \cos \theta_X \cos \theta_{\bar X'}) \,.
\label{ec:Cdist}
\end{equation}
The angles $\theta_X$, $\theta_{\bar X'}$ are measured using as spin axis
the parent top (anti)quark momentum in the $t \bar t$ CM system. The factor
\begin{equation}
C \equiv \frac{\sigma(t_R \bar t_R) + \sigma(t_L \bar t_L) -
\sigma(t_R \bar t_L) - \sigma(t_L \bar t_R)}{\sigma(t_R \bar t_R) +
\sigma(t_L \bar t_L) + \sigma(t_R \bar t_L) + \sigma(t_L \bar t_R)}
\label{ec:C}
\end{equation}
is the relative number of like helicity minus opposite helicity $t \bar t$
pairs, and measures the spin correlation between the top quark and
antiquark. 
It is also interesting to study the relative distribution of one spin
analyser from the $t$ quark and other from the $\bar t$. Let $\varphi_{X \bar
X'}$ be the angle between the three-momentum of $X$ (in the $t$ rest frame) and 
of $\bar X'$ (in the $\bar t$ rest frame). The angular distribution can be
written as~\cite{Bernreuther:2004jv}
\begin{equation}
\frac{1}{\sigma} \frac{d\sigma}{d\!\cos \varphi_{X \bar X'}} =
\frac{1}{2} (1+D \, \alpha_X \alpha_{\bar X'} \cos \varphi_{X \bar X'}) \,,
\label{ec:Ddist}
\end{equation}
with $D$ a constant defined by this equality. The actual values of $C$ and $D$
depend to some extent on the parton distribution functions (PDFs) used and the $Q^2$ scale at which they are evaluated, and also on the invariant mass of the $t \bar t$ pair $m_{t \bar t}$. At the tree level $C \simeq 0.314$,
$D \simeq -0.212$ while at one loop $C = 0.326 \pm 0.012$,
$D = -0.237 \pm 0.07$~\cite{Bernreuther:2004jv}. (See also Ref.~\cite{Bernreuther:2010ny}.) An upper cut on $m_{t \bar t}$ increases these values. 

Using the spin analysers $X$, $\bar X'$ for the respective decays of $t$, $\bar
t$, one can define the asymmetries
\begin{eqnarray}
A_{X \bar X'} & = & \frac{N(\cos \theta_X \cos \theta_{\bar X'} > 0) 
- N(\cos \theta_X \cos \theta_{\bar X'} < 0)}
{N(\cos \theta_X \cos \theta_{\bar X'} > 0) +
N(\cos \theta_X \cos \theta_{\bar X'} <0)} \,, \notag \\
\tilde A_{X \bar X'} & = & \frac{N(\cos \varphi_{X \bar X'} > 0) 
- N(\cos \varphi_{X \bar X'} < 0)}{N(\cos \varphi_{X \bar X'} > 0) +
N(\cos \varphi_{X \bar X'} < 0)}  \,,
\end{eqnarray}
whose theoretical values derived from Eqs.~(\ref{ec:Cdist}) and (\ref{ec:Ddist})
are
\begin{eqnarray}
A_{X \bar X'} & = & \frac{1}{4} C \alpha_X \alpha_{\bar X'} \,, \notag \\
\tilde A_{X \bar X'} & = & \frac{1}{2} D \alpha_X \alpha_{\bar X'} \,.
\label{ec:Ath}
\end{eqnarray}
As we have shown in section~\ref{sec:5}, $\alpha_{\bar X} = - \alpha_X$ in full generality, so that for charge conjugate decay channels we have
$\alpha_{X'} \alpha_{\bar X} = \alpha_X \alpha_{\bar X'}$ and the asymmetries
$A_{X' \bar X} = A_{X \bar X'}$, $\tilde A_{X' \bar X} = \tilde A_{X \bar X'}$
are equivalent.

\begin{figure}[htb]
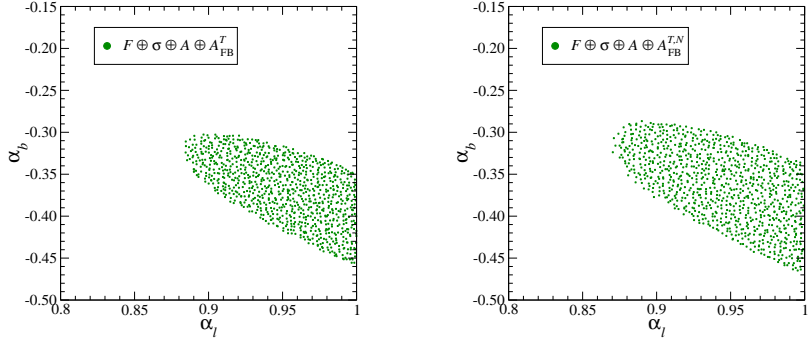

\begin{center}
\begin{tabular}{ccc}
\epsfig{file=Figs/fit-R-alab.eps,height=4.5cm,clip=} & \quad \quad &
\epsfig{file=Figs/fit-C-alab.eps,height=4.5cm,clip=}
\end{tabular}
\caption{Variation of $\alpha_\ell$ and $\alpha_b$ for anomalous couplings within the $1\sigma$ limits in Fig.~\ref{fig:fitR} (real case, left) and Fig.~\ref{fig:fitC} (complex case, right).}
\label{fig:fit-a}
\end{center}
\end{figure}

The expected LHC precision in the measurement of the asymmetries $\all$, $\allt$ (dropping superscripts to easy the notation) in the dilepton channel has been estimated in Ref.~\cite{Hubaut:2005er}, using an invariant mass cut $m_{t \bar t} < 550$ GeV to enhance the spin correlation, 
\begin{align}
& \all = -0.101 \pm 0.005\;\text{(stat)} \pm 0.006\;\text{(sys)} \, \notag \\ 
& \allt = 0.145 \pm 0.0055\;\text{(stat)} \pm 0.005\;\text{(sys)} \,.
\end{align}
Assuming that no new physics contributes to the top decay, these measurements can be directly translated into measurements of $C$ and $D$, by setting $\alpha_{\ell^+} = 1$ and using Eqs.~(\ref{ec:Ath}),
\begin{align}
& C = 0.404 \pm 0.020\;\text{(stat)} \pm 0.024\;\text{(sys)} \,, \notag \\
& D = -0.290 \pm 0.011\;\text{(stat)} \pm 0.010\;\text{(sys)} \,.
\end{align}
On the other hand, if we allow for new physics in the decay then $\alpha_{\ell^+}$ can significantly deviate from unity. We present in Fig.~\ref{fig:fit-a} the variation of $\alpha_{\ell^+}$ and $\alpha_b$ for anomalous couplings within the $1\sigma$ regions of Fig.~\ref{fig:fitR} (real $Wtb$ vertex) and Fig.~\ref{fig:fitC} (complex).
As we can observe, the assumption $\alpha_{\ell^+} = 1$, although legitimate if one wants to concentrate on new physics in $t \bar t$ production, does not hold if new physics is allowed in the decay. In this case, the possible presence of anomalous couplings introduces an additional uncertainty in the extraction of $C$ and $D$ from $\all$ and $\allt$,
\begin{align}
& C = 0.404 \pm 0.020\;\text{(stat)} \pm 0.024\;\text{(sys)} ~^{+0.129}_{-0} \;(\alpha) \,, \notag \\
& D = -0.290 \pm 0.011\;\text{(stat)} \pm 0.010\;\text{(sys)} ~^{+0}_{-0.103}\;(\alpha) \,,
\end{align}
which is much larger than the experimental one. In any case,
since $|\alpha_{\ell^+}| \leq 1$, absolute values of $C$, $D$ larger than the SM ones would
do not suffer from this uncertainty. Hence, if deviations in this direction were measured, they would clearly point to new physics in the production. For other top spin analysers the dependence on anomalous couplings is more pronounced (see Figs.~\ref{fig:a} and \ref{fig:fit-a}) and the associated uncertainties in the extraction of $C$ and $D$ larger.

\section{Signals of new physics in top decays}
\label{sec:8}

A model-independent determination of the $Wtb$ vertex, such as the one discussed in section~\ref{sec:6}, must not rely on any assumption regarding the nature or size of the possible anomalous $Wtb$ couplings. In particular, all of them must be let completely arbitrary, with all cancellations among their contributions allowed. Were it not for these cancellations, the combination of single top cross sections and $W$ helicity fractions~\cite{Chen:2005vr} would suffice to obtain good limits on anomalous couplings.

On the other hand, in definite SM extensions it may well happen that not all anomalous couplings are different from zero. Then, it makes sense to investigate the (discovery) limits when only one of them is non-zero. With this purpose, we use the most sensitive single observables: the ratio $\rp = \fp/\fz$ and the asymmetry $\Ap$ measured in $t \bar t$ production, with the sensitivities estimated in Ref.~\cite{AguilarSaavedra:2007rs} (corresponding to 10 fb$^{-1}$ at 14 TeV) and the central values corresponding to the SM prediction,
\begin{align}
& \rp = 0.00051 \pm 0.0021\;\text{(stat)} \pm 0.0016\;\text{(sys)} \,, \notag \\
& \Ap = 0.5472 \pm 0.0032\;\text{(stat)} \pm 0.0099\;\text{(sys)} \,.
\end{align}
In addition, we use $\afbN$ in $t$-channel single top production, with the precision given in Eq.~(\ref{ec:Ameas2}).
The $3\sigma$ discovery limits on anomalous couplings, taking $\vl=1$ and assuming that only one of them is non-zero at a time, are presented in Table~\ref{tab:lim}. We also assume for simplicity that couplings are either real or purely imaginary. For completeness we also include the limits on $\vl$, although if we assume that no other physics is present the limits from $t$-channel single top production are much better~\cite{Aad:2009wy,Ball:2007zza,AguilarSaavedra:2010wf}. In each case, the observable yielding the $3\sigma$ deviation is also indicated.
On the right column we include the $3\sigma$ discovery limits from $b \to s \gamma$. They are obtained using~\cite{Grzadkowski:2008mf}
\begin{equation}
10^4 \times \mathrm{Br}(b \to s \gamma) = (3.15 \pm 0.23) -8.2 \, (\vl-1) + 427 \, \vr - 712 \, \gl + 1.9 \, \gr + \dots
\label{ec:bsgamma}
\end{equation}
and the experimental value $\mathrm{Br}(b \to s \gamma) = (3.52 \pm 0.23 \pm +0.09) \times 10^{-4}$~\cite{Barberio:2008fa}. For the real part, limits are directly obtained from Eq.~(\ref{ec:bsgamma}) while the limits on imaginary parts involve quadratic terms, estimated from calculations in Refs.~\cite{Grzadkowski:2008mf,Larios:1999au}. 
\begin{table}[t]
\begin{center}
\begin{tabular}{ccc}
\multicolumn{2}{c}{Top observables} & $b \to s\gamma$ \\
\hline \\[-4mm]
\begin{tabular}{l}
$\RE \, \vl \leq 0.62$ \\ $\RE \, \vl \geq 1.21$
\end{tabular}
& ($\sigma_{tW}$) &
\begin{tabular}{l}
$\RE \, \vl \leq 0.83$ \\ $\RE \, \vl \geq 1.07$
\end{tabular}
\\[6mm]
\begin{tabular}{l}
$\RE \, \vr \leq -0.111$ \\ $\RE \, \vr \geq 0.18$
\end{tabular}
& ($\rp$) &
\begin{tabular}{l}
$\RE \, \vr \leq -0.0015$ \\ $\RE \, \vr \geq 0.0032$
\end{tabular}
\\[5mm]
$|\IM \, \vr| \geq  0.14$ & ($\rp$) &
$|\IM \, \vr| \gtrsim  0.01$
\\[2mm]
\begin{tabular}{l}
$\RE \, \gl \leq -0.083$ \\ $\RE \, \gl \geq 0.051$
\end{tabular}
& ($\rp$) &
\begin{tabular}{l}
$\RE \, \gl \leq -0.0019$ \\ $\RE \, \gl \geq 0.00090$
\end{tabular}
\\[5mm]
$|\IM \, \gl| \geq  0.065$ & ($\rp$) &
$|\IM \, \gl| \gtrsim  0.006$
\\[2mm]
$|\RE \, \gr| \geq 0.056$ & ($\Ap$) &
\begin{tabular}{l}
$\RE \, \gr \leq -0.33$ \\ $\RE \, \gr \geq 0.76$
\end{tabular}
\\[5mm]
$|\IM \, \gr| \geq  0.115$ & ($\afbN$) & --
\end{tabular}
\end{center}
\caption{Estimated $3\sigma$ discovery limits on anomalous couplings, assuming that only one of them is non-zero at a time and that they are either real or purely imaginary. The coupling $\vl$ is real by definition.}
\label{tab:lim}
\end{table}
Several important comments regarding these results are in order:
\begin{itemize}
\item[(i)] The good sensitivity of $\mathrm{Br}(b \to s \gamma)$ to $\vr$ and $\gl$ makes it unlikely to obtain a positive signal from them in top decays, unless some other new physics cancels their contribution to the former (a possibility which is not excluded).
\item[(ii)] Conversely, top decay observables are much more sensitive to $\gr$, either real or not. This fact makes their experimental study quite interesting, since it is expected that $|\gr| \gg |\vr|,|\gl|$ in SM extensions~\cite{Bernreuther:2008us}.
\item[(iii)] For top decay observables, the effect of the $b$ quark mass in the limits on $\RE \vr$ and $\RE \gl$ is quite sizeable. If the $b$ quark mass was neglected these limits would be symmetric, and numerically equal to the limits on $\IM \vr$ and $\IM \gl$, respectively.
\item[(iv)] The $3\sigma$ discovery limits from top decay observables are numerically much smaller than the $1\sigma$ model-independent ones in section~\ref{sec:6}. This fact clearly shows that cancellations are still at work in those.
\end{itemize}

\section{Summary}
\label{sec:9}

In this work we have investigated the decay of a polarised top quark into a polarised $W$ boson and a massive $b$ quark, using the most general $Wtb$ vertex arising from dimension-six gauge invariant effective operators. Our starting point has been the calculation of the spin density matrix for this decay in the $W$ helicity basis. This matrix contains eight dimensionless form factors which are functions of the $Wtb$ couplings. Of these, three are determined by the total top width and the $W$ helicity fractions, measurable (in principle) in the decay of unpolarised top quarks.

For the decay of polarised top quarks we have defined and calculated the transverse and normal polarisation fractions. They are analogous to the helicity fractions but using the directions transverse and normal to the $W$ momentum, as depicted in Fig.~\ref{fig:axes}. These quantities are very useful to access two of the off-diagonal form factors in the density matrix. In particular, the normal polarisation fractions are sensitive to complex phases in the anomalous couplings. We have introduced and calculated a forward-backward asymmetry
$\afbN$ which vanishes in the SM and for real anomalous couplings and is very sensitive to the phase of the coupling $\gr$ in the Lagrangian of Eq.~(\ref{ec:lagr}),
\begin{equation}
\afbN \simeq - 0.64 \, P \, \IM \vl \gr^* \,.
\end{equation}
This asymmetry (or related ones~\cite{Kane:1991bg}) would be easy to measure, since it only involves the decay products of either a top or antitop quark, in contrast with
more complicated asymmetries in $t \bar t$ production, built using triple products with momenta from both $t$ and $\bar t$ decays. Moreover, the latter also rely on the spin correlation between $t$ and $\bar t$, while $\afbN$ directly probes the imaginary terms in the density matrix (being for this reason more sensitive). Its measurement in $t$-channel single top production is expected to have a good accuracy, based on previous estimations for similar observables.

We have obtained a sum rule relating the $\fTz$, $\fNz$ components with the helicity fractions $\Fpm$ and shown that, given present and expected limits on the latter, the remaining components $\fTpm$, $\fNpm$ may significantly deviate from the SM prediction.
Thus, their measurement at LHC is necessary and will bring new information about the $Wtb$ vertex. For $\fTpm$, this information is related to the one obtained from spin asymmetries in the top quark rest frame, and these measurements are expected to be complementary. For $\fNpm$, the information on complex phases of anomalous couplings is a novel ingredient, independent from other sources.

Using all the relevant observables, namely $W$ polarisation fractions (or related quantities), asymmetries in the top quark rest frame and the $tW$ cross section, we have performed a fit to the general (complex) $Wtb$ vertex, taking the single top polarisation as a free parameter. The code {\tt TopFit\,2} has been implemented and used for this purpose. The most interesting result here, rather than the precise values of the limits obtained, is the fact that limits on all $Wtb$ couplings can actually be obtained in a model-independent way for the general complex case, and that the single top polarisation can be cleanly extracted from measurements as well. These results, new in the literature, are non-trivial because of the many possible cancellations among contributions from anomalous couplings, which require the inclusion of top polarisation-related observables (in addition to helicity fractions and cross sections) to be reduced.
As a by-product of this fit, the allowed variation of the spin analysing power constants of top decay products has been obtained. This result has been used to estimate the ``theoretical'' uncertainty in the measurement of top-antitop spin correlations in $t \bar t$ production, associated to the possibility of new physics in the top decay.

The counterpart of the model-independent determination of the $Wtb$ vertex is the sensitivity to ``single'' anomalous couplings, assuming that only one of them is non-zero. This analysis is relevant from the theoretical point of view, because in definite SM extensions it seems quite possible that not all anomalous $Wtb$ couplings will be simultaneously different from zero. In this case ({\em i.e.} without cancellations) we have seen that $3\sigma$ deviations in selected observables will be possible for anomalous couplings at the $0.05-0.1$ level. In particular, for the anomalous coupling $\gr$, which is expected to be the largest in SM extensions, the sensitivity of top decay angular asymmetries is an order of magnitude better than from the $b \to s \gamma$ branching ratio. A detailed study of top decay observables at Tevatron and LHC is hence compulsory. Finally, we note that
$3\sigma$ discovery limits are found to be numerically smaller than the $1\sigma$ model-independent ones, due to cancellations still present. This fact motivates further investigation of observables sensitive to the three remaining form factors in the spin density matrix. That work is left for future studies.

\section*{Acknowledgements}

We thank M. P\'erez-Victoria, J. Prades and J. Santiago for interesting discussions.
This work has been partially supported by projects FPA2006-05294 and FPA 2008-02878 (MICINN), FQM 101 and FQM 437 (Junta de Andaluc\'{\i}a), PROMETEO 2008/004 (Generalitat Valenciana), CERN/FP/83588/2008 (FCT),
and by the European Community's Marie-Curie Research Training
Network under contract MRTN-CT-2006-035505 ``Tools and Precision
Calculations for Physics Discoveries at Colliders''. The work of J.A.A.S. has been supported by a MICINN Ram\'on y Cajal contract.

\appendix
\section{$W$ boson polarisation vectors}

The gauge boson polarisation vectors in the helicity basis  are well known. Taking the
 positive $z$ axis in the direction of the $W$ boson momentum in the top quark rest frame, they are
\begin{align}
& \varepsilon_z^0 = \frac{1}{M_W} (q,0,0,E_W) \,, \notag \\
& \varepsilon_z^+ = - \frac{1}{\sqrt 2} (0,1,i,0) \,, \notag \\
& \varepsilon_z^- = \frac{1}{\sqrt 2} (0,1,-i,0) \,.
\end{align}
The vectors corresponding to the transverse basis can be simply obtained by performing a $90^\circ$ rotation around the $x$ axis (see Fig.~\ref{fig:axes}). They are
\begin{align}
& \varepsilon_y^0 = \frac{i}{\sqrt 2} (\varepsilon_z^+ + \varepsilon_z^-)
 = (0,0,1,0) \,, \notag \\
& \varepsilon_y^+ = \frac{1}{2} (\varepsilon_z^+ - \varepsilon_z^-) + \frac{i}{\sqrt 2} \varepsilon_z^0
 =  - \frac{1}{\sqrt 2} (\frac{-i q}{M_W},1,0, \frac{-i E_W}{M_W}) \,, \notag \\
& \varepsilon_y^- = - \frac{1}{2} (\varepsilon_z^+ - \varepsilon_z^-) + \frac{i}{\sqrt 2} \varepsilon_z^0
 = \frac{1}{\sqrt 2} (\frac{i q}{M_W},1,0,\frac{i E_W}{M_W}) \,.
\end{align}
Analogously, the vectors for the normal basis are obtained with a rotation around the $y$ axis,
\begin{align}
& \varepsilon_x^0 = - \frac{1}{\sqrt 2} (\varepsilon_z^+ - \varepsilon_z^-)
 = (0,1,0,0) \,, \notag \\
& \varepsilon_x^+ = \frac{1}{2} (\varepsilon_z^+ + \varepsilon_z^-) + \frac{1}{\sqrt 2} \varepsilon_z^0
 = \frac{1}{\sqrt 2} (\frac{q}{M_W},0,-i, \frac{E_W}{M_W}) \,, \notag \\
& \varepsilon_x^- = \frac{1}{2} (\varepsilon_z^+ + \varepsilon_z^-) - \frac{1}{\sqrt 2} \varepsilon_z^0
 = - \frac{1}{\sqrt 2} (\frac{q}{M_W}0,i,\frac{E_W}{M_W}) \,.
\end{align}

\end{document}